\pdfoutput=1
\PassOptionsToPackage{table}{xcolor}
\documentclass[]{elsarticle}
\usepackage[a4paper, total={6in, 8in}]{geometry}

\usepackage{graphicx}
\usepackage{etoolbox}
\usepackage{comment}
\usepackage{color}
\usepackage{tikz}
\usetikzlibrary{calc,shapes,arrows,shapes.geometric,fit,matrix,backgrounds,plotmarks,patterns}
\usepackage{pgfplots}
\usepackage{pgfplotstable}
\usepackage{framed}
\usepackage{balance} 
\usepackage{soul}
\usepackage[normalem]{ulem}	
\usepackage{multirow}
\usepackage{colortbl}
\usepackage{cancel}
\usepackage{lineno}
\usepackage{algorithm}
\usepackage{algorithmicx}
\usepackage[noend]{algpseudocode}
\usepackage{amsmath,amsfonts,amssymb,mathtools}
\usepackage{wasysym}

\newtheorem{myex}{Example}

\newtheorem{mydef}{Def.}
\newtheorem{myprob}{Problem}

\newtheorem{mythm}{Theorem}
\newtheorem{myfact}{Fact}

\newtheorem{mylemma}{Lemma}
\newtheorem{myrmk}{Remark}
\newtheorem{myproof}{Proof}

\DeclarePairedDelimiter\ceil{\lceil}{\rceil}

\definecolor{dartmouthgreen}{rgb}{0.05, 0.5, 0.06}
\definecolor{viridian}{rgb}{0.25, 0.51, 0.43}
\definecolor{wheat}{rgb}{0.96, 0.87, 0.7}
\definecolor{wenge}{rgb}{0.39, 0.33, 0.32}
\definecolor{ameth}{rgb}{0.6, 0.4, 0.8}
\makeatletter
\newcommand{\vastt}{\bBigg@{0.9}}
\newcommand{\vast}{\bBigg@{3.35}}
\newcommand{\Vast}{\bBigg@{5.5}}
\newcommand{\reg}{$\text{\emph{reg}}$}
\makeatother

\makeatletter
\def\dynscriptsize{\check@mathfonts\fontsize{\sf@size}{\z@}\selectfont}
\makeatother
\def\textunderset#1#2{\leavevmode
  \vtop{\offinterlineskip\halign{%
    \hfil##\hfil\cr\strut#2\cr\noalign{\kern-.1ex}
    \hidewidth\dynscriptsize\strut#1\hidewidth\cr}}}

\usepackage{subcaption}
\DeclareCaptionLabelSeparator{periodspace}{.\quad}
\usepackage{enumitem}
\setlist{leftmargin=5.5mm}
\usepackage[linktocpage=true,breaklinks]{hyperref}

\usepackage{lineno}

\begin{document}

\pgfplotstableread[col sep=comma]{./data/prior0.csv}\priorzero
\pgfplotstableread[col sep=comma]{./data/priorspecial.csv}\priorspecial
\pgfplotstableread[col sep=comma]{./data/prior1.csv}\priorone
\pgfplotstableread[col sep=comma]{./data/prior2.csv}\priortwo
\pgfplotstableread[col sep=comma]{./data/prior3.csv}\priorthree
\pgfplotstableread[col sep=comma]{./data/prior4.csv}\priorfour
\pgfplotstableread[col sep=comma]{./data/prior5.csv}\priorfive
\pgfplotstableread[col sep=comma]{./data/prior6.csv}\priorsix
\pgfplotstableread[col sep=comma]{./data/step.csv}\stepf


\setlength{\footskip}{60pt}


\title{Bsmooth: Learning from user feedback to disambiguate query terms in interactive data retrieval}


\author[ibm]{Bernardo Gon\c{c}alves\corref{cor1}}
\ead{bng@br.ibm.com}
\author[umich]{H. V. Jagadish}
\ead{jag@umich.edu}
\address[ibm]{IBM Research, S\~ao Paulo, Brazil}
\address[umich]{University of Michigan, Ann Arbor, USA}

\cortext[cor1]{Corresponding author: IBM Research; Rua Tut\'oia 1157, S\~ao Paulo, SP 04007-900, Brazil.}

\begin{abstract}
There is great interest in supporting imprecise queries (e.g., keyword search or natural language queries) over databases today. To support such queries, the database system is typically required to disambiguate parts of the user-specified query against the database, using whatever resources are intrinsically available to it (the database schema, data values distributions, natural language models etc).  
Often, systems will also have a user-interaction log available, which can serve as an extrinsic resource to supplement their model based on their own intrinsic resources.  
This leads to a problem of how best to combine the system's prior ranking with insight derived from the user-interaction log. Statistical inference techniques such as maximum likelihood or Bayesian updates from a subjective prior turn out not to apply in a straightforward way due to possible noise from user search behavior and to encoding biases endemic to the system's models. 
In this paper, we address such learning problem in interactive data retrieval, with specific focus on type classification for user-specified query terms. We develop a novel Bayesian smoothing algorithm, \textsf{Bsmooth}, which is simple, fast, flexible and accurate. We analytically establish some desirable properties and show, through experiments against an independent benchmark, that the addition of such a learning layer performs much better than standard methods. 
\end{abstract}


\journal{arXiv.CS}

\begin{frontmatter}
\end{frontmatter}


\section{Introduction}\label{sec:intro}
\noindent
The last decade has seen significant research activity on imprecise query processing over structured data. Non-expert end-users are accessing databases, often over the web, and posing queries as keyword collections, natural language sentences, fielded values with some partial structure, and even as clicks on a graphical user interface (GUI). The system then bears the responsibility of disambiguating the given imprecise query (e.g., a keyword search query, or a natural language question) onto a complete, precise query (e.g., in SQL) that best reflects the user's intent. 

The typical state-of-the-art approach to the problem is based on scoring functions that synthesize various resources such as the database schema, data values distributions, natural language models etc, which are all intrinsic to the database. If user feedback can be captured from previous sessions, it should be possible to use it as an extrinsic resource to learn from and improve upon the system's intrinsic interpretation, as we see in this example.

\begin{myex}\label{ex:hanks}
Consider query Q, alternatively specified in keyword and natural language against a relational store of the Internet Movie Database (IMDb) in its original design. 

Q. \emph{\underline{`tom hanks'} 2004}.\\\indent
Q$\,^\prime$. \emph{find all movies played by \underline{`tom hanks'} in 2004}.

\vspace{5pt}
\noindent
For term \emph{`tom hanks,'} this is the top-5 ranking of disambiguation options (in IMDb's own \textsf{TABLE.attribute} design) according to some scoring function intrinsic to the database.\footnote{Details of this scoring function are not important for this paper. Suffice to say that the values are obtained from the real system described in \cite{li2015}.}

\noindent
\begin{small}
\begin{tabular}{c l l} 
\indent 1. & \textsf{CHAR\_NAME.name} & (27.93\%)\\
\indent 2. & \textsf{NAME.name} & (23.46\%)\\
\indent 3. & \textsf{TITLE.title} & (22.07\%)\\
\indent 4. & \textsf{MOVIE\_INFO.info} & (15.08\%)\\
\indent 5. & \textsf{ROLE\_TYPE.role} & (11.45\%) 
\end{tabular}
\end{small}

The system is unsure about whether `tom hanks' is the name of a character, name of a person, or a movie title.
The desired answer is \textsf{NAME.name}, which is ranked second. Based on information intrinsic to the database including a language model, this is the best the system has been able to do.  $\Box$ 
\end{myex}

\begin{figure}[t]
\includegraphics[width=1.03\textwidth]{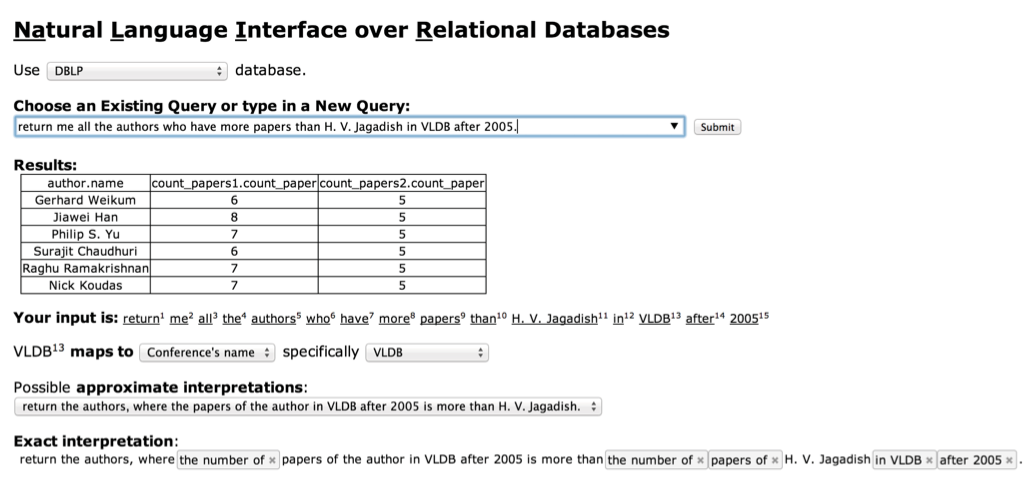}
\caption{Example of graphical user interface (GUI) where users can provide implicit feedback by selecting a better interpretation of an ambiguous query term for classification against the database schema types (source: \cite{li2015}). The challenge is how to best exploit (learn from) such feedback, which is accumulated over query sessions like this one.}\label{fig:nalir}
\end{figure}

By learning from user feedback provided interactively over some sessions, we consider whether the system in this example might have been able to do better.  
Specificallly, we consider the following scenario:
Query answer takes place in two steps: (S1) the imprecise query is 
disambiguated token-wise onto existing database elements, and then (S2) a precise SQL query with possible join operations is generated and processed to return with the answer. The session is optionally interactive after S1 to allow the user to correct the interpretation of the query, before it is actually rendered in SQL and evaluated in S2. 
A possible GUI embodying such scenario (over DBLP, a different example database than IMDb) is shown in Fig.~\ref{fig:nalir}.

\setcounter{myex}{0}
\begin{myex}\label{ex:hanks-cont} (continued). 
Now, for the same term \emph{`tom hanks'} previously discussed, this is an instance of 
a simulated user feedback log collected by simulation from a crowdsourced experiment that we describe later. 
Such an information source is considered extrinsic to the database.\vspace{-5pt}\\ 

\noindent
\begin{small}
\begin{tabular}{c l l} 
\indent 1. & \textsf{CHAR\_NAME.name} & (1)\\
\indent 2. & \textsf{NAME.name} & (8)\\
\indent 3. & \textsf{TITLE.title} & (0)\\
\indent 4. & \textsf{MOVIE\_INFO.info} & (1)\\
\indent 5. & \textsf{ROLE\_TYPE.role} & (0) 
\end{tabular}
\end{small}
 
\noindent
We find that 8 out of 10 users in this interaction log tend to think \emph{`tom hanks'} is the name of a person, but the remaining 2 users prefer other choices.
Our problem is to learn from this user feedback. More precisely, combine it with the intrinsic system scores shown above that suggest character name is the most likely interpretation. $\Box$
\end{myex}

Our inference problem seems to suit bayesian learning: the intrinsic model is a prior distribution, and every user choice is an observation that can be used to update that distribution.
However, we will see that user feedback may not always be as high-quality as in the case of the example above, thus it is not clear how best to combine the rankings from the two sources.  At the very least, we have to choose how to weight these two sources of information.  There is no reason to believe that the two are equally important or equally likely. 
There is much good work on combining evidence from multiple sources. However, the incomplete and unreliable nature of the two sources are not similar. They have structural asymmetries that we introduce next, which will render related work techniques (discussed in depth in \S\ref{sec:related-work}) less appropriate for such an interactive data retrieval use case. 

\textbf{Implicit feedback}. 
Explicit feedback 
may be too burdensome to ask of the user.  A system cannot usually count on users taking the trouble to correct its mistakes, unless such corrections are on their critical path to completing their task of interest. Therefore, user logs with explicit feedback are likely to be small in size and coverage, if they exist at all.
However, observed user behavior after the query can provide implicit feedback: 
if a user opts to change the default (top-ranked) option and try another ranked match for a query term we get implicit feedback into an interaction log; but if a user does nothing, it could indicate that she is satisfied (so-called `good abandonment' \cite{tokuda2009}) or that she is frustrated (bad abandonment).

\textbf{User-feedback noise}. 
Natural language words may have some genuine ambiguity. Moreover user search behavior may have many oddities: we must admit a range of patterns from 
hesitant decisions and exploratory choices that may well not be consistent with one another. 
Given a sufficiently large user feedback log, one can hope that such noise will be suppressed.  However, the population of queried terms often form a `long tail' so that many terms have only limited feedback and hence are highly susceptible.
 
\textbf{DB-encoding noise}. 
The specific DB encoding at both schema and data levels highly influences the system scoring of disambiguation choices. In IMDb, e.g., person's names are phrased in a more formal manner (e.g., `Hanks, Tom'), while names of movie characters are phrased in a more informal manner (e.g., character `Tom Hanks', played in comedy `Bamboo Shark') just like users tend to do in a query. As spelling similarity is one useful measure for matching terms (proper names in particular), the intrinsic scoring is biased against matching `tom hanks' to the person's name. At the schema level, likewise, IMDb's table `NAME' (which stores records of people) also obscures the disambiguation of a query like \textsf{person hanks}, since the word `person' does not appear in the table or attribute name. 
Were such encoding bias not present, the DB-intrinsic scoring would be more stable and possibly amenable for characterization as a more reliable subjective prior. 
In view of this bias, the scorings (e.g., as in Example~\ref{ex:hanks}), while having indicative value, are not good probability estimates. Therefore, it is not obvious how to use the computed scores in a principled manner for composition with another information source.

\textbf{Research problem}. 
Given a query term comprising one or more words (e.g.,`tom hanks'), our task is to determine the type of the term.  To perform this term type disambiguation, we have available to us two 
noise-prone information sources.  For queries comprising multiple terms, we assume that an upstream process correctly tokenizes the query and then we solve the type disambiguation independently for each token (term). 


We devise a novel bayesian smoothing solution to this problem, where we use empirically observed noise in the data source to determine its relative weight while accounting for the various expected biases that we discussed above.  When user feedback is unambiguously focused on a single answer, as in Example~\ref{ex:hanks-cont}, our solution clearly converges to this answer, almost irrespective of the intrinsic prior. Yet it gives much more credence to the prior when user feedback is more mixed. 

\textbf{Contributions}. In short, the contributions of this paper are as follows. 
We present a bayesian model and 
an algorithm, \textsf{Bsmooth}, that uses the up-to-date rankings (coming from the two sources) and compensates for the amount of noise found. 
We also describe how to tune \textsf{Bsmooth}'s parameters given an arbitrary DB-intrinsic scoring profile --- so that it can be incorporated by any competing Interactive Data Retrieval (IDR) system with its own characteristic scoring function. We evaluate the effectiveness of \textsf{Bsmooth} for IDR against an independent and carefully designed benchmark \cite{coffman2014}. 
We also compare it with 9 competitive systems reported in the relational keyword search literature. 
We report results from explicit and implicit user interactions simulated by crowdsourcing for data collection, and then shed light on their comparison. 

The paper is organized as follows. 
In \S\ref{sec:settings} we present all the data we have collected for this study. In \S\ref{sec:entropy} we review Shannon's entropy, and introduce it as an uncertainty measure that can be used promptly to analyze our collected data. In \S\ref{sec:learning} we present the \textsf{Bsmooth} model and algorithm. In \S\ref{sec:eval} we evaluate its effectiveness against the benchmark. In \S\ref{sec:implicit} we analyze implicit feedback, comparing with explicit feedback. Insights here may be useful for other uncertainty management applications. In 
\S\ref{sec:related-work}  we present an in-depth discussion of related work, and in \S\ref{sec:conclusions} conclude the paper.

\section{Preparatory Data Collection}\label{sec:settings}
\noindent 
Six related data sets, which we name D0 through D5, are central to describing our work, and to its empirical evaluation.
In this section we introduce our notation and these datasets, preparing for our formal problem set up. 

\textbf{Database and Benchmark}\label{subsec:dataset}.
D0 is the base dataset, from which everything else is derived.  We require that there be a set B of benchmark queries against D0, with known answers.
While the specific choice of B and D0 is immaterial for our conceptual development, we will need to specify these for our empirical evaluation.
To make matters more concrete for the reader, and to be able to present examples with ease, we reveal our choices of B and D0 here upfront.
Note that these specific choices only matter for the purpose of the evaluation.

For B, we use an independent benchmark developed to evaluate relational keyword search systems in the literature \cite{coffman2010,coffman2014}. It follows the standards of the Text Retrieval Conference (TREC) series, 
and provides binary relevance assessments on answers to a set of 50 imprecise queries. So it provides B, where $|\text{B}|=50$, and does it separately for each of some popular datasets that have been used by empirical evaluations in related work, e.g., IMDb, Wikipedia, and DBLP.  

For D0, we use IMDb (release 2009), because of its well-known semantic domain of movies, which fits the scenario of an average user on the web issuing search queries over a relational database whose actual design is unknown to her --- recall IMDb's \textsf{TABLE.attribute} design from Example~\ref{ex:hanks}. 
As of April 26, 2017, it is available for download as a PostgreSQL dump.\footnote{\url{http://www.cs.virginia.edu/~jmc7tp/resources.php}.} 
We have loaded and used it as is. It is the relational database we will disambiguate search queries against.

The 50 numbered search queries in B, when issued against IMDb, give rise to a total of 62 query terms. 
The remaining 5 data sets each have scores for each possible type match for each of these 62 query terms.
As an aid to memory, we optionally add a character string after each data set number.  Thus, we say D1-INTR or D1 to mean the same thing.

\textbf{D1-INTR. (DB-intrinsic source)}. 
We first ran the benchmark queries on our system \cite{li2015} to get the DB-intrinsic scoring (and induced ranking) for each of the 62 query terms. Example~\ref{ex:hanks} is based on such scoring for query term `tom hanks.' Overall, this dataset D1-INTR has 60\% precision-at-rank-one (P@1) accuracy, as we will see in detail later in \S\ref{sec:eval}.

\begin{figure*}[t]
\begin{subfigure}{.49\textwidth}
\begin{framed}
\vspace{-3pt}
\begin{scriptsize}
\noindent
\texttt{In this query on a movie database:}\vspace{0.5pt} 

\texttt{title \underline{indiana jones}}\vspace{3pt}\\ 
\colorbox{cyan!40}{?} $\,$\texttt{Guess what the underlined term means}:
\vspace{3pt}\\
\begin{tabular}{lll}
$\!\!\APLbox\!\!$ & \texttt{MOVIE\_INFO.info} & \framebox{\phantom{aaaaaaaaaaaaaaaaaaaaaaaaa}}\\
$\!\!\APLbox\!\!$ & \texttt{TITLE.title} & \framebox{\vspace{-4pt}\phantom{aaaaaaaaaaaaaaaaaaaaaaaaa}\vspace{-4pt}}\\
$\!\!\APLbox\!\!$ & \texttt{NAME.name} & \framebox{\phantom{aaaaaaaaaaaaaaaaaaaaaaaaa}}\\
$\!\!\APLbox\!\!$ & \texttt{CHAR\_NAME.name} & \framebox{\phantom{aaaaaaaaaaaaaaaaaaaaaaaaa}}\\
$\!\!\APLbox\!\!$ & \texttt{ROLE\_TYPE.role} & \framebox{\phantom{aaaaaaaaaaaaaaaaaaaaaaaaa}}
\end{tabular}\vspace{3pt}\\
\colorbox{red!40}{!} $\,$\texttt{Required}:\\
\texttt{1.$\!$ Check one or more best matches to the term}.\vspace{-2pt}\\
\texttt{2.$\!$ Describe what you think each option means}.
\end{scriptsize}
\vspace{-3pt}
\end{framed}
\end{subfigure}
\begin{subfigure}{.49\textwidth}
\begin{framed}
\vspace{-3pt}
\begin{scriptsize}
\noindent
\texttt{In this query on a movie database:}\vspace{0.5pt} 

\texttt{title \underline{indiana jones}}\vspace{5pt}\\ 
\noindent
\colorbox{cyan!40}{?} $\,$\texttt{Guess what the underlined term means}:
\vspace{6pt}\\
\begin{tabular}{ll}
$\!\!\APLbox\!\!$ & \texttt{MOVIE\_INFO.info}\vspace{2pt}\\
$\!\!\APLbox\!\!$ & \texttt{TITLE.title}\vspace{2pt}\\
$\!\!\APLbox\!\!$ & \texttt{NAME.name}\vspace{2pt}\\
$\!\!\text{\rlap{$\checkmark$}}\square\!\!$ & \texttt{CHAR\_NAME.name}\vspace{2pt}\\
$\!\!\APLbox\!\!$ & \texttt{ROLE\_TYPE.role}\\
\end{tabular}\vspace{8.3pt}\\
\colorbox{red!40}{!} $\,$\texttt{Required}:\\
\texttt{1.$\!$ Choose the best match to the term}.\vspace{-9pt}\\
\end{scriptsize}
\vspace{-3pt}
\end{framed}
\end{subfigure}
\vspace{-1pt}
\caption{Crowd task designs to collect explicit feedback (left) and an abstraction of implicit (right) user feedback in a cost-effective simulation of a real system. The right-hand side task is implicit feedback when the user just accepts an option marked by default, and it simulates the GUI shown in Fig.~\ref{fig:nalir} also in the case when the user changes the option given as default.}
\label{fig:crowd-tasks}
\vspace{-8pt}
\end{figure*}

Then we used crowdsourcing to simulate explicit and implicit user feedback as a data collection task, preceding any modeling carried out by us.\footnote{In this paper we do not refer to `crowdsourcing models' designed to post-process (filter) workers' answers. We only use simple crowdsourcing for experimental data collection based on best practices reported in \cite{kittur2008}.} 
In fact, crowdsourcing has recently been pointed out to be a cost-effective experimentation paradigm to simulate user interactions in Interactive Information Retrieval \cite{zuccon2013}. 
We designed two crowd tasks (Fig. \ref{fig:crowd-tasks}) and deployed them at the \textsf{Microworkers.com} platform, paying \$0.10 USD per task instance. 
The first experiment simulates explicit feedback.

\textbf{D2-EXPL. (Explicit user feedback)}. 
D2 stores the observed counts for the candidate matches as explicitly stated by crowd workers representing typical users. 
For each of the 62 query terms, 
10 different workers have been recruited.\footnote{Although a sample size of 10 may look small, it is more realistic as most queried terms may fall in the log's `long tail,' having limited feedback; besides, we will see later in \S\ref{sec:eval} that it has been enough to successfully disambiguate most terms, so it is hard to see how a larger sample could improve our settings.}  
In each task instance we ask the worker what the best matches (one or more) are for the term. 
For example, we refer to query Q22 in the benchmark set B to reproduce here our crowd task design, see Fig. \ref{fig:crowd-tasks} (left). 
Regarding this design: 
\begin{itemize}
\item The term is presented in the context of its query to simulate the system scenario more realistically; 

\item The crowd workers are reasonably assumed to be unfamiliar with the IMDb internal schema types, i.e., have no knowledge of the database structural details;

\item The order of the IMDb types is assigned randomly in each instance of the task (seen by one worker). 

\item 
To push workers to process the task cognitively, we require them to describe their choices briefly; their descriptions, being unstructured, are not useful to our results compilation --- only the options they check; 
\end{itemize}

\textbf{D3-IMPL. (Implicit user feedback)}. 
In D2, we required user feedback for every task.  
Now in a second crowd experiment (Fig. \ref{fig:crowd-tasks}, right), 
we give users the option to just accept the system-provided default and provide no feedback.
For each of the 62 terms the task is re-instantiated to 10 workers with no participation in the first experiment. This task design has all properties listed above except that two new properties hold:

\begin{itemize}
\item A default choice is given. We pick it from the highest-scored option (top-1st) according to dataset D1-INTR, which as mentioned has 60\% P@1 and is available to us upfront; e.g., for query term `indiana jones' the default is \textsf{CHAR\_NAME.name}, for `gone with the wind' such a default is \textsf{TITLE.title}; 

\item It is single-choice, while the task design for explicit feedback is multi-choice. This is to reproduce an optional user interaction, where in fact only one option can be chosen at a time. 
\end{itemize}

\noindent
So this second experiment is a simulation of implicit feedback. 
D3 stores the observed counts for candidate matches abstracted to be implicitly indicated by users for each of the query terms. 

\textbf{D4-RAND (Random feedback)}. 
For each query term, we have sampled from the uniform distribution an integer $0 \leq c \leq 10$ as counts for each disambiguation option. 
Thereby, we obtain a dataset that reproduces the scenario of a noisy 
interaction log.  
D4 will be useful to test how proposed methods behave in the presence of clearly noisy data --- a scenario that has to be accounted for in a real system that takes user feedback (with its possible oddities) as input and changes its behavior in response.

D4 stores the observed counts for candidate matches that results from random user input.  This really poor result contrasts with D5, described next.

\textbf{D5-BENCH. (Benchmark's relevance assessments)}. 
D5 is the golden answer key, with the known correct matches for each query term.  
For each of the 50 imprecise queries in B, the benchmark informs which tuples in the database (D0) are considered relevant --- as a binary classification, so each tuple is either relevant or not.\footnote{In adherence to the Cranfield paradigm \cite{cleverdon1997}, TREC does not distinguish between highly relevant and partially relevant results (cf. \cite[p.~6]{coffman2010}).}  

From such tuple-level classification we derive an attribute-level classification as follows. Consider, e.g., query Q7 --- which has a single term `tom hanks' --- and its relevant tuple(s) R7.
\vspace{4pt}

Q7. `tom hanks'\vspace{6pt}\\
\indent
R7.  \hspace{0pt}
\begin{tabular}{c|c|c}
NAME & id & name\\ \hline & 393050 & Hanks, Tom\\
\end{tabular}

\noindent
Now recall the 5 options to disambiguate `tom hanks' from Example~\ref{ex:hanks} (rendered by intrinsic D1, by the way). 
From R7, we know straightforwardly that the one relevant match for the term is \textsf{NAME.name}. We do the same for all of the 62 terms. A highlight is that, when converted from tuple to attribute level, the relevance assessments turn out to be all singletons, i.e., each term has exactly one relevant type to disambiguate it, which will facilitate our evaluation in \S\ref{sec:eval}. Query (term) Q14, e.g., has (R14) 3 tuples assessed relevant all of which `aggregate' to \textsf{TITLE.title}. 

\vspace{3pt}
Q14. `lord of the rings'\vspace{6pt}\\
\indent
R14. \hspace{0pt}
\begin{tabular}{c|c| c |c}
$\!\!\!$TITLE$\!\!$ & id & title & $\!\!$production\_year$\!\!$\\
\hline
& $\!\!$513253$\!$ & $\!$The Lord of the Rings: $\!\!$...$\!\!$ & 2003\\
& $\!\!$513256$\!$ & $\!$The Lord of the Rings: $\!\!$...$\!\!$ & 2002\\
& $\!\!$513250$\!$ & $\!$The Lord of the Rings: $\!\!$...$\!\!$ & 2001\\
\end{tabular}\vspace{3pt}\\

\noindent
This is how dataset D5 is rendered, gathering the specific IMDb types that are the golden answer key matches to the query terms in the benchmark.

\vspace{-3pt}
\section{Shannon's Entropy as Uncertainty Measure}\label{sec:entropy}
\noindent
Since the scores obtained from the intrinsic and extrinsic sources are derived independently, using completely different techniques, an important question to ask is how potentially correct each source is for a particular term. Of course, without knowing the right answer, our system has no way to judge absolute correctness on the fly.  However, it can judge how certain any source is, using classic entropy measures. Intuitively, we could rely more upon a source that is more certain, in a manner that we will make mathematically precise in the next section.

Given a \emph{scoring vector} $\vec{\theta}\!=\!(\theta_1, \theta_2, ..., \theta_k)$ where $\theta_i \geq 0 $ for all $1 \leq i \leq k$ and $\sum_i^k \theta_i =1$, that is, $\vec{\theta}$ is a probability distribution, Shannon's entropy (Eq.~\ref{eq:shannon}) gives a measure of the expected surprise (uncertainty) in it \cite{murphy2012}. 
\vspace{-1pt}
\begin{eqnarray}
\label{eq:shannon-entropy}
H_b(\vec{\theta}) \;=\; - \textstyle\sum_i^k{ \theta_i\; \log_b\,\theta_i} 
\vspace{-3pt}
\end{eqnarray}
Base $b$ defines a particular unit, which for $b = 2$ is bits. Entropy is a bounded function. For vector $\vec{\theta}$ with $k > 0$ possibilities, it is minimum and maximum (resp.) when $\vec{\theta}\!=\!(1, 0, ..., 0)$ and $\vec{\theta}\!=\!(1/k, 1/k, ..., 1/k)$. 
Entropy will measure how undecided vector $\vec{\theta}$ is. 
For any $k$, we have $0 < H_b(\vec{\theta}) \leq \log_b{k}$. 
So we can plug Eq.~\ref{eq:shannon} to normalize entropy (with no reference to a specific base $b$) in the $[0, 1] \subset \mathbb R$ interval. Our measure is now unified for any $k \in \mathbb N$ and base $b$ for having a comparable threshold. 
\vspace{-2pt}
\begin{eqnarray}
\label{eq:shannon}
H(\vec{\theta}) = H_b(\vec{\theta})/\log_b{k}
\vspace{-3pt}
\end{eqnarray}

We can also easily convert counts from an observation vector $\vec{n}\!=\!(n_1,\, ...,\, n_k)$, where $n_i \in \mathbb N$, to a scoring vector like $\vec{\theta}$ above using relative frequency $\theta_i := n_i / \sum_i^k n_i$. If we view the DB-intrinsic scoring as a prior hypothesis and the DB-extrinsic counts (whether D2 or D3) as an observation, we can use $H(\vec{\theta})$ to measure how `regular' evidence data is towards the most frequently observed outcome in the sample space. 
In Example~\ref{ex:hanks-cont} (contd.), which is derived from D3-IMPL, we have 8 counts for \textsf{NAME.name} out of 10 for term `tom hanks.' Its counts vector has $H(\theta) \approx 0.397$, for normalized $0 < H \leq 1$. 
When the distribution is very skewed, like in this case, we will have low entropy --- meaning strong evidence towards the most favored outcome. 

We define \emph{ambiguity threshold} to be an entropy value that 
distinguishes when evidence is ``good'' from when it is ``bad''.\footnote{Evidence being ``bad'' does not mean that its probability distribution is wrong, as some terms may be genuinely ambiguous; e.g., `terminator' may well refer to either a character name or a movie title. However, it may still mean that we should be less likely to trust high-entropy counts vectors as opposed to low-entropy ones. This intuition will be empirically evaluated in \S\ref{subsec:accuracy}.} 
The precise value we choose for this threshold will turn out to be a tuning parameter that we will explore empirically in \S\ref{sec:eval}. 
At this point, we only need the intuition that a data source with entropy less than the threshold provides solid evidence while one with entropy greater than the threshold is more ambiguous about the evidence it provides.

We close this section with some empirical observations regarding the observed entropies in our data sets.
Recall that the D1-INTR is already in the form of scoring vectors. 
For D2-D4 we apply relative frequency 
to give them the scoring vector form as well. Then, for each of these datasets and each of the 62 terms, we compute $H(\theta)$ by Eq.~\ref{eq:shannon}. Table~\ref{tab:entropy} shows the entropy mean $\bar{H}$ and standard deviation (stdev) that we have found. 
Some observations about Table~\ref{tab:entropy} are:

\begin{table}[t]
\caption{Entropy mean $\bar{H}$ and standard deviation (stdev).} 
\label{tab:entropy}
\begingroup\setlength{\fboxsep}{0.25pt}
\colorbox{gray!9}{
\begin{tabular}{c | c} 
Dataset$\!\!$ & Entropy mean $\bar{H} \pm$ stdev.$\!\!$\\ 
\hline
$\!\!$D1-INTR & $98.47 \pm 01.83\%$\\ 
$\!$D2-EXPL & $48.13 \pm 18.32\%$\\ 
$\!\!$D3-IMPL & $19.85 \pm 28.07\%$\\ 
D4-RAND & $88.63 \pm 01.34\%$\\ 
\end{tabular}
}\endgroup
\end{table}

\begin{itemize}
\item Every IDR system will have its own intrinsic scoring function, which as we have discussed is prone to DB-encoding noise. Our particular D1-INTR has very high entropy mean with very low stdev, which indicates that entropy is nearly the same over queried terms and may not help to indicate (in)accuracy for D1-INTR;

\item D2-EXPL has higher entropy mean than D3-IMPL, while for stdev the inverse holds. This can be easily explained by the workers being able to choose multiple choices per task in Fig.~\ref{fig:crowd-tasks} (left), but only one choice per task in Fig.~\ref{fig:crowd-tasks} (right). In fact D2-D3 have been collected for different purposes: (D2) a closer to ``true'' disambiguation distribution as a baseline, and (D3) a more realistic interaction log; 

\item For both D2-EXPL and D3-IMPL, the entropy mean, even with added stdev, is far below D4-RAND's entropy mean, which means that the feedback provided by a crowd of users can indeed be informative. 
\end{itemize}

\noindent
The first observation above suggests that the intrinsic source may not be sensitive to entropy, while the two other observations suggest that the extrinsic source is. We will take that into account in the analytical study that comes next, and revisit this assumption later in our evaluation in \S\ref{sec:eval}.

\section{The Bayesian Smoothing Model}\label{sec:learning}
\noindent
In this section we define relevant concepts and state the inference problem in light of the general bayesian smoothing model.\footnote{At this point, one may consider a review on relevant bayesian inference concepts that is given in Appendix~\ref{subsec:tutorial}.} We then consider progressively more involved solutions, ultimately arriving at our proposed algorithm, \textsf{Bsmooth}.

\begin{mydef}\label{def:scoring}
Let $k \in \mathbb N^\ast$. We say that $\vec{x}$ is a \textbf{scoring vector} of size $k$, and write $\vec{x}(x_1, ..., x_k)$, if we have $x_i \geq 0$ for all bins $i \in \mathbb N^\ast$ where $i \leq k$ and $\sum_i^k{x_i=1}$. 
\end{mydef}

\begin{mydef}\label{def:convex}
Let $\vec{x}$ and $\vec{y}$ be two scoring vectors. We say that $\vec{z}=(z_1, ..., z_k)$ is a \textbf{convex combination of} $\vec{x}$ and $\vec{y}$, and write $\vec{z} = \vec{x} \oplus \vec{y}$, if there exist $\omega_1, \omega_2 \geq 0$ with $\omega_1+ \omega_2=1$ such that $z_i= \omega_1\, x_i + \omega_2\, y_i$ for all bins $i \in \mathbb N^\ast$, $i \leq k$.
\end{mydef}

Bayesian inference allows us to arrive at a posterior combining information from two sources --- if we take the DB-intrinsic scoring as a prior, which is updated with the DB-extrinsic observations to get a posterior. As the prior and posterior scoring vectors take values in the continuous parameter space, bayesian inference takes the form of \emph{bayesian smoothing} \cite{murphy2012} (cf. Appendix~\ref{subsec:tutorial}). Since for each instance of disambiguation problem we have (with no loss of generality) a number of $k > 2$ types as options to disambiguate, the model will take the form of a \emph{Dirichlet-multinomial} \cite[p.~78]{murphy2012}. 

\begin{myrmk}\label{rmk:applied-problem}
The extent to which we can consider the DB-extrinsic counts as `observations' in our interactive data retrieval problem (i.e., as empirical counts, even if subject to some smoothing) is still not quite clear. We will proceed from here as if so, to gain depth into the abstract problem, and then revisit this assumption later in \S\ref{subsec:accuracy}. 
$\Box$
\end{myrmk}

\subsection{The Bayesian Smoothing Problem}\label{subsec:problem}
\noindent
Bayesian smoothing in terms of the dirichlet-multinomial can be obtained from the \emph{dirichlet posterior mean} as given by Eq.~\ref{eq:dm}, where $p(\theta_i \,|\, \vec{n})$ is the posterior probability of type $i \in \mathbb N^\ast,\, i \leq k$ as a disambiguation option; and $n_i$, $\alpha_i$ are components of the observed counts vector $\vec{n}$ and hyper-parameter vector $\vec{\alpha}$ with $n,\, \alpha$ as the total observations and the hyper-prior concentration parameter. 
\begin{eqnarray}\label{eq:dm}
p(\theta_i \,|\, \vec{n}) \!\!&=&\!\! \frac{n_i + \alpha_i}{n+\alpha}
\end{eqnarray}

Here, $\vec{n}$ maps naturally to our observation counts on each type $i$, while $\vec{\alpha}$ maps to our DB-intrinsic scoring vector as a \mbox{(hyper-)}prior. The formula simply gives their weighted sum based on the scale of $\alpha \in \mathbb R^+$ against $n \in \mathbb N$. The open problem is how to set this relative weight, i.e., the strength of the prior, or alternatively, the weight of evidence. What we really have here is a convex combination problem where the scaling of $\alpha$ must be done at query time.

\begin{myprob}\label{prob:scale}
$\!$Let $\vec{x}(x_1, ..., x_k)\!$ be a scoring vector rendered by the DB-intrinsic source, and $\vec{n}(n_1, ..., n_k)$ with $\sum_i^k n_i=n$ be the counts vector supplied by the DB-extrinsic source, both at query time. For notation and analytical convenience, the counts vectors $\vec{n}$ is normalized into a scoring vector $\vec{y}$. 

So we have two scoring vectors $\vec{x}$ and $\vec{y}$ from (resp.) the intrinsic and extrinsic sources and want to find \mbox{$\vec{z} = \vec{x} \oplus \vec{y}$} that gives the best results for all queried terms. That is, we seek $\omega_1, \omega_2 \geq 0$ with $\omega_1 \!+\! \omega_2=1$ such that $z_i \!= \omega_1 x_i + \omega_2 y_i$ for all $i \leq k$, $i \in \mathbb N^\ast$, such that the correctness of $\vec{z}$ (which we will measure as P@1), averaged over queried terms, is maximum. We call $\omega_2$ the \textbf{weight of evidence}.  
$\Box$
\end{myprob}

We can just set $\alpha := n\, (\omega_1/\omega_2)$ to scale $\vec{\alpha}$ accordingly and get the posterior scoring vector $\vec{\theta}\in \mathbb R^k$ estimated by bayesian smoothing within the $k$-simplex continuous parameter space. This is bayesian smoothing where the (hyper\-)prior has an empirical basis.

\subsection{Baseline Schemes: The Limiting Cases}
\noindent
A first and very simple baseline method to consider is Maximum Likelihood Estimation (MLE), which corresponds to bayesian smoothing with $\omega_2=1$ in Problem~\ref{prob:scale}.  

\begin{myrmk}\label{rmk:mle}
Simply set $\omega_2=1$ $(\omega_1=0)$. 
For reference we call this method \emph{\textbf{MLE}}.
$\Box$
\end{myrmk}
MLE considers only the observations, and ignores any prior obtained from an intrinsic model. It can work very well if there are enough observations and they are low entropy. However, such an assumption is too strong. MLE seems too naive and we do not expect it to perform well in general. 

The opposite limiting case is to set $\omega_2=0$ $(\omega_1=1)$; that is, to ignore all observations and just stay with the intrinsic model irrespective of the observed implicit user feedback.  This is trivial to do, and obviously misses the whole point of improving disambiguation by learning from user behavior.  

A smarter technique would choose between these two limiting cases based on which one is more likely to be correct.

\begin{myrmk}\label{rmk:activate}
Given an ambiguity threshold $D \in [0, 1]$, the weight $\omega_2 \in [0, 1]$ could be defined simply as $\omega_2=1$ ($\omega_1=0$) if $H(\vec{y}) \leq D$, and $\omega_2=0$ ($\omega_1=1$) otherwise. This would be a (discontinuous) staircase function we call \textbf{\emph{STEP}}. 
 $\Box$
\end{myrmk}

\subsection{Balanced Convex Combinations}\label{subsec:convex}
\noindent
In this subsection, we study balance properties of convex combinations, so as to be able to design less extreme weighting functions.

\begin{myex}\label{ex:convex}
Given two scoring vectors $\vec{x}=(1, 0, 0, 0, 0)$ and $\vec{y}=(1/3, 1/3, 0, 1/3, 0)$, one can ask, say, is $\vec{z}=(1/2,$ $1/4, 0, 1/4, 0)$ a convex combination of $\vec{x}$ and $\vec{y}$? Yes, we can pick $\omega_1=1/4$ and $\omega_2=3/4$ so that $\vec{z}$ can be written $z_i=(1/4)\, x_i + (3/4)\, y_i$ for all bins $1 \leq i \leq 5$. $\Box$
\end{myex}

\begin{mylemma}\label{lemma:convex}
Let $\vec{x}$ and $\vec{y}$ be two scoring vectors and $t \neq u$ two of their bins. Then a convex combination $\vec{z}=\vec{x} \oplus \vec{y}$ with $z_t=z_u$ exists and is unique if and only if one of these mutually exclusive conditions hold:\vspace{1pt}\\
\indent \emph{(i)}$\;\;$ $(x_t - x_u)/(y_t - y_u) < 0$,\vspace{2pt}\\
\indent \emph{(ii)}$\;$ $(x_t - x_u) = 0$ but $(y_t - y_u) \neq 0$,\vspace{2pt}\\
\indent \emph{(iii)} $(y_t - y_u) = 0$ but $(x_t - x_u) \neq 0$. 
\end{mylemma}

\begin{myproof}
We show the statement by construction of an abstract convex combination $\vec{z}$ satisfying $z_t=z_u$. The conditions stated by the lemma are then tied up to the existence of a unique pair $\omega_1, \omega_2 \geq 0$ as the solution set of a linear system in these two variables. See Appendix~\ref{proof:lemma-convex}. $\blacksquare$
\end{myproof} 

Note that of the three cases in the lemma, only the first is a ``true'' convex combination: the other two are really corner cases where we ignore one of the two inputs.

\setcounter{myex}{1}
\begin{myex} (contd.)
For the vectors $\vec{x}=(1, 0, 0, 0, 0)$ and $\vec{y}=(1/3, 1/3, 0, 1/3, 0)$, suppose that we wanted to find any convex combination $\vec{z}=\vec{x}\,\oplus\,\vec{y}$ with $z_2=z_4$. The one that has been presented satisfies it --- recall $\,\vec{z}\,=(1/2, 1/4, 0,$ $1/4, 0)$ rendered out of $\omega_1=1/4,\, \omega_2=3/4$. Yet there are infinitely many pairs $\omega_1, \omega_2 \geq 0$ with $\omega_1 + \omega_2=1$ that we could have picked so that $z_2=z_4$. In fact, by Lemma~\ref{lemma:convex} we should not expect uniqueness here as we have both $(x_2-x_4)=(y_2-y_4)=0$.  In contrast, if we want $z_1 \!=\! z_3$, that is not possible because $x_1 > x_3$ and $y_1 > y_3$.  $\Box$
\end{myex}

\begin{mydef}\label{def:top-bin}
Let $\vec{x}=(x_1, ..., x_k)$ be a scoring vector and $1 \leq t \leq k$ one of its bins. We say that $t$ is a \textbf{top bin of $\vec{x}$}, and write $t \in \ceil*{\vec{x}}$, if $x_t \geq x_i$ holds for all bins $1 \leq i \leq k$. 
\end{mydef}
\begin{mydef}\label{def:disagreed-top-bins}
Let $\vec{x}=(x_1, ..., x_k)$ and $\vec{y}=(y_1, ..., y_k)$ be two scoring vectors with top bins $t \in \ceil*{\vec{x}}$ and $u \in \ceil*{\vec{y}}$. We say that $t$ and $u$ are \textbf{disagreed top bins} (of $\vec{x}$ and $\vec{y}$), and write $t \asymp u$, if either $t \notin \ceil*{\vec{y}}$ or $u \notin \ceil*{\vec{x}}$ or both.  
\end{mydef}
\begin{mythm}\label{thm:convex}
Let $\vec{x}=(x_1, ..., x_k)$ and $\vec{y}=(y_1, ..., y_k)$ be two scoring vectors with top bins $t \in \ceil*{\vec{x}}$ and $u \in \ceil*{\vec{y}}$. If these are disagreed top bins $t \asymp u$, then a convex combination $\vec{z}=\vec{x} \oplus \vec{y}$ with both $z_t=z_u$ exists and is unique. 
\end{mythm}
\begin{myproof}
From Lemma~\ref{lemma:convex} and Def.~\ref{def:top-bin}--\ref{def:disagreed-top-bins}, see Appendix~\ref{proof:thm-convex}. $\blacksquare$
\end{myproof}

Note that when the intrinsic and extrinsic vectors $\vec{x}$ and $\vec{y}$ agree on their top bins, there is little for us to do.  The interesting case is when they have disagreed top bins, and this is precisely the case covered by Theorem~\ref{thm:convex}.  The theorem states that we can solve a linear system in two variables, $\omega_1, \omega_2 \geq 0$ to find a balanced combination where two top choices are equally highly scored.  Intuitively, this balance point is important because the final top choice answer tends to be decided one way or the other whether we increase $\omega_1$ or $\omega_2$.

\begin{figure}[t]
\advance\leftskip 3.0 cm
\pgfplotsset{every axis legend/.append style={ at={(0.31,0.535)},
anchor=south east}}
\begin{tikzpicture}[y=2cm, x=2cm,font=\sffamily,scale=0.92]
    \begin{axis}[
        xmajorgrids = true,
        ymajorgrids = true,
        y label style = { yshift=-0.1cm },
        x label style = { yshift=0.01cm },
        xtick={0,0.1,...,1.1},
        ytick={0,0.1,...,1.1},
        height=7.525cm,
        width=8.5cm,
        xmin=0,
        xmax=1,
        ymin=0,
        ymax=1,
        xlabel={\textsf{entropy of evidence $H(\vec{y}_j)$}},
        ylabel={\textsf{weighting threshold $\bar{\omega}_2$}},
]
\addplot[color=black, style=thick, mark=square*, mark size=1.0pt]   table [x=h, y=w, col sep=comma] {\priorzero};
\addlegendentry{\scriptsize{PRIOR $\vec{x}_6$}}
\addplot[color=blue, solid, mark=triangle*, mark size=1.5pt]  table [x=h, y=w, col sep=comma] {\priorone};
\addlegendentry{\scriptsize{PRIOR $\vec{x}_5$}}
\addplot[color=green!50!gray, solid, mark=*, mark size=1.2pt]  table [x=h, y=w, col sep=comma] {\priortwo};
\addlegendentry{\scriptsize{PRIOR $\vec{x}_4$}}
\addplot[color=red, solid, mark=square*, mark size=1.0pt]  table [x=h, y=w, col sep=comma] {\priorthree};
\addlegendentry{\scriptsize{PRIOR $\vec{x}_3$}}
\addplot[color=orange, solid, mark=triangle*, mark size=1.5pt]  table [x=h, y=w, col sep=comma] {\priorfour};
\addlegendentry{\scriptsize{PRIOR $\vec{x}_2$}}
\addplot[color=olive, densely dotted, mark=*, mark size=1.2pt]  table [x=h, y=w, col sep=comma] {\priorspecial};
\addlegendentry{\scriptsize{PRIOR $\vec{x}_1$}}
\addplot[color=violet, solid, mark=*, mark size=1.2pt, mark options={fill=red}]  table [x=h, y=w, col sep=comma] {\priorfive};
\addlegendentry{\scriptsize{PRIOR $\vec{x}_0$}}
\addplot +[color=black!100, mark=none, dashed] coordinates {(0.75, 0) (0.75, 1)};
\addplot +[color=black!100, mark=none, dashed] coordinates {(0, 0.203) (1, 0.203)};
\addplot +[color=black!100, mark=none, dashed] coordinates {(0, 0.0425531915) (1, 0.0425531915)};
     \end{axis}
\end{tikzpicture}
\caption{Thresholds to balance out disagreed top bins for a class of priors of fixed skewness. Legend keys follow the order the curves appear in the vertical axis. The more skewed the prior the higher lies its curve. For $\vec{y}(.7875, .2125)$, e.g., which has $H(\vec{y})\approx.75$, one needs only $\omega_2 \approx .0035$ to balance $\vec{x}_0$. }
\label{fig:priors}
\end{figure}
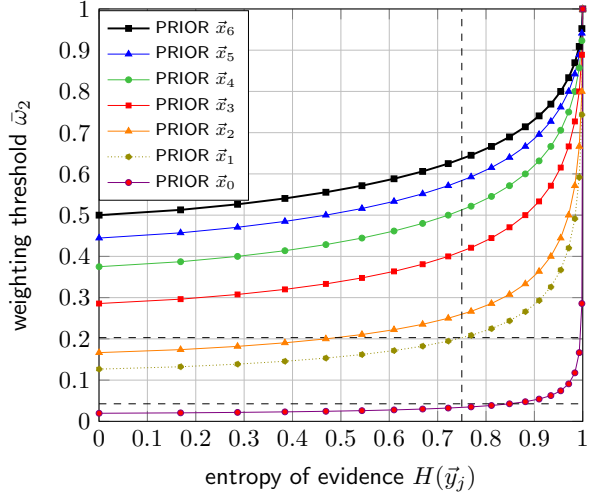

To develop a better intuition for this balance point, 
we consider a wide range of values for $\vec{x}$ and $\vec{y}$ as shown in Table~\ref{tab:priors}. The goal of using such a range of values is to explore balance conditions in terms of the varying entropy measures $H(\vec{y})$, as given by Eq.~\ref{eq:shannon}.  
To avoid degenerate cases, we use $(.499,.501)$ rather than $(.5,.5)$, and $(.001,.999)$ rather than $(0,1)$.
We consider every entry in the first row paired with every entry in the second row.
For each pair $\vec{x_i}, \vec{y_j}$, where $\vec{y_j}$ has its own entropy measure, we solve the linear system in variables $\bar{\omega}_1, \bar{\omega}_2 \geq 0$.
Fig.~\ref{fig:priors} shows the results. 

The higher is the entropy, the higher is the required threshold $\bar{\omega}_2$ for balance, except for the least skewed $\vec{x}_0$ (violet) whose $\bar{\omega}_2$ is nearly constant most of the domain.

\subsection{Baseline Scheme: Linear Weighting}\label{subsec:continuous}

\begin{myex}\label{ex:consensus}
Let $\vec{x}=\!(.45, .43, .12, 0, 0)$ and $\vec{y}=\!(.08, .45,$ $.47, 0, 0)$ be two scoring vectors. Note that we have different sets of top bins $\ceil*{\vec{x}}=\{1\}$ and $\ceil*{\vec{y}}=\{3\}$. 
Then by Theorem~\ref{thm:convex} we know that a unique convex combination $\vec{z}=\vec{x} \oplus \vec{y}$ exists with $z_1=z_3$. In fact, this is $z_i \approx .54\,x_i + .46\,y_i$ so that $\vec{z} \approx (.28, .44, .28, 0, 0)$. Now we have $\ceil*{\vec{z}}=\{2\}$, which is not one of the top bins of $\vec{x}$ nor of $\vec{y}$. 
\footnote{This is really not a problem as this ``consensus'' top bin $\ceil*{\vec{z}}\!\!=\!\!\{2\}$ is brought forth naturally out of balancing the two vectors, and not by an inferential scheme that pushes it through individually at the bin granularity level --- cf. related discussion in \ref{sec:related-work} of Dempster-Shafer's theory as applied in related work.} 

Entropy is $H(\vec{y})\approx .569$, which is not very high. So if the ambiguity threshold is, say, $D=.75$, then we have $H(\vec{y}) < D$. Now recall the baseline method of STEP from Remark~\ref{rmk:activate}. The top bin picked would be $\ceil*{\vec{z}}=\{3\}$, just as if out of MLE from $\vec{y}$. 
$\Box$
\end{myex}

\begin{table*}[t!]
\caption{Two separate rows: entry $\vec{x_i}$ in first row applies to each entry $\vec{y_i}$ in second row, rendering one curve in Fig.~\ref{fig:priors}. That is, first row shows the prior skew levels from $\vec{x_0}$ to $\vec{x_6}$. Second row shows the several points (entropy levels) $\vec{y_0}$ to $\vec{y_{20}}$. }
\label{tab:priors}
\advance\leftskip -0.15cm
\begin{footnotesize}
\begingroup\setlength{\fboxsep}{0.25pt}
\colorbox{gray!15}{%
\begin{tabular}{c | c | c | c | c | c | c | c}
$\!\!$PRIOR$\!\!$ & $\vec{x}_0(.499, .501)\!\!$ & $\!\vec{x}_1(.4275, .5725)\!\!$ & $\!\vec{x}_2(.4, .6)\!\!$ & $\!\vec{x}_3(.3, .7)\!\!$ & $\!\vec{x}_4(.2, .8)\!\!$ & $\!\vec{x}_5(.1, .9)\!\!$ & $\!\vec{x}_6(.001, .999)\!\!$\\
\hline
\hline
$\!\!$POINT$\!\!$ & $\!\vec{y}_0(.999, .001)\!\!$ & $\!\vec{y}_1(.975, .025)\!\!$ & $\!\vec{y}_2(.950, .050)\!\!$ &  $\hdots$ & $\!y_{18}(.550, .450)\!\!$ & $\!y_{19}(.525, .475)\!\!$ & $\!y_{20}(.501, .499)\!\!$\\
\end{tabular}
}\endgroup
\end{footnotesize}
\end{table*}

We see that STEP has an undesirable discontinuity jump: a very small change in the extrinsic scoring's entropy $H(\vec{y})$, say, a single user clicking one thing differently, say in the process of exploring her disambiguation options, can completely change the system output.  Instead, we would like that such small changes should not add 
too large a change in its weight $\omega_2$. At the very least, we would like 
$\omega_2\!: [0, 1] \to \mathbb R$ to be a continuously differentiable function: its first derivative exists all over $[0, 1]$ and is itself a continuous function.

\begin{myrmk}\label{rmk:linear}
Given the entropy $H(\vec{y})$ of extrinsic scoring $\vec{y}$, its weight $\omega_2\!: [0, 1] \to [0, 1]$ could be defined as a linear function $\omega_2(H)=\beta_0 - \beta\,H$ for some constants $\beta_0,\, \beta > 0$. Note the negative sign (with $\beta > 0$) to meet the intuition of inverse correlation: the higher is $H(\vec{y})$, the less likely is $\vec{y}$ to be accurate. 
Now if we also take boundary conditions $\omega_2(0)=1$ and $\omega_2(1)=0$ as two related constraints, then we easily infer $\beta_0=\beta=1$ and get more specific $\omega_2(H)=1 - H$. We call this method \textbf{\emph{LINEAR}}. 
 $\Box$
\end{myrmk}

\subsection{Proposed Scheme: Log-linear Weighting}\label{subsec:logit}
\noindent
By Example~\ref{ex:linear} we illustrate why a linear weighting profile cannot be adequate in general. 
\begin{myex}\label{ex:linear}
Let us revisit Example~\ref{ex:consensus} with $\vec{x}=\!(.45, .43,$ $.12, 0, 0)$, but now with $\vec{y}^{\,\prime}\!=\!(.08, .35,$ $.37, .1, .1)$ instead. By Lemma \ref{lemma:convex} the unique convex combination that balances out top bins $\ceil*{\vec{x}}=\{1\}$ and $\ceil*{\vec{y}^{\,\prime}}=\{3\}$ exists and is precisely $z_i \approx .468\,x_i + .532\,y_i^{\,\prime}$ so that $\vec{z} \,\approx (.253, .387, .253, .053,$ $.053)$. Again we have $\!\ceil*{\vec{z}} = \{2\}$, not a top bin of $\vec{x}$ nor of $\vec{y}^{\,\prime}$. 

But here we have entropy $H(\vec{y}^{\,\prime}) \approx .885$, which is fairly high. The same ambiguity threshold, $D=.75$, would now make $H(\vec{y}^{\,\prime}) > D$. Thus we have: by MLE, $\ceil*{\vec{z}}=\{3\}$; by STEP, $\ceil*{\vec{z}}=\{1\}$; and by LINEAR, $\omega_1\approx .885,\, \omega_2 \approx .115$, then $\vec{z}\approx\!(.407, .421, .149, .012, .012)$ and therefore $\ceil*{\vec{z}}=\{2\}$. 
$\Box$
\end{myex}

It is not surprising that LINEAR fails to bend to penalize entropy properly in the neighborhood of the ambiguity threshold $H=D \pm \delta$, since it has a constant rate of change w.r.t. entropy, $\omega^\prime_2(H)=-1$. 
Furthermore, for any chosen ambiguity threshold, $D$, we would like the weight to be set as
 $\omega_2(D)=\bar{\omega_2}$ plotted in Fig.~\ref{fig:priors} for each prior. Recall the reason for this crucial constraint from \S\ref{subsec:convex}. At the ambiguity turning point $D$ we do \emph{not} know which scoring to favor, the extrinsic or the intrinsic. 
LINEAR --- being a straight line --- obviously cannot fit all constraints in $P\!=\! \{\,(0, 1), (1, 0),$ $(D, \bar{\omega}_2)\,\}$ with third point $(D,\, \bar{\omega}_2)$ virtually anywhere in the plane $[0, 1]^2$. 

To address this shortcoming, we desire a function that can fit the extreme points $(0, 1), (1, 0)$ but still have room to fit the third point $(D,\, \bar{\omega}_2)$. 
Furthermore, consider that our extrinsic scoring on a given term is either very skewed or very entropic, (resp.) near complete certainty or complete uncertainty. At such states, we want to change the weight of evidence only slowly, with some lag, say, to respond to one or another user with exploratory behavior. We want to implement inverse correlation (cf. Remark~\ref{rmk:linear}) linearly otherwise. This is achieved with the \emph{logit transform} \cite{shalizi2015}.

\begin{myrmk}\label{rmk:logit}
The logit transform is given by $\textsf{logit}(\theta)=\ln{[\theta/(1-\theta)]}$, and is also called log-odds. 
We model the relationship between $\omega_2$ and $H$ to be log-odds linear, as given by Eq.~\ref{eq:logit} for constants $\beta_0,\beta>0$.  
\begin{eqnarray}\label{eq:logit}
\ln{\frac{\omega_2}{1-\omega_2}} = \beta_0 - \beta\,H
\end{eqnarray}
The logit is the inverse of the logistic function: let both sides be exponents of base $e$ and easily solve for $\omega_2$ to get Eq.~\ref{eq:logistic}. 
\begin{eqnarray}\label{eq:logistic}
\omega_2(H) = \frac{1}{1 + e^{-(\beta_0-\beta\,H)}}
\end{eqnarray}
We call this method \emph{\textbf{LOGIT}}. 
$\Box$
\end{myrmk}

\noindent
We assess LOGIT against any linear function in Fact~\ref{fact:logit}. 

\begin{myfact}\label{fact:logit}
Let $P=\{(0, 1),\, (1, 0),\, (D, \bar{\omega}_2)\}$ be a set of points in the plane $[0, 1]^2$, where $\bar{\omega}_2$ is taken over each of the weighting thresholds considered --- acquired out of Theorem~\ref{thm:convex} given the DB-intrinsic profiles shown in Table~\ref{tab:priors}. To instantiate this fact, we will keep the same $D=.75$ we have been using in previous examples, and revisit the tuning of this parameter in \S\ref{subsec:accuracy}. 

$\!\!\!$Now let the lack of fit for a function $f\!\!: [0, 1] \to \mathbb R$ be defined by least squares $L_f=\sum_{(a, b)\in P}{(\,b-f(a)\,)^2}$; and $f_1(x)=\beta_0-\beta\,x$ stand for linear functions (e.g., LINEAR) whereas $f_2(x)=1/(1+e^{-(\beta_0-\beta\,x)})$ stand for LOGIT over $\beta_0, \beta >0$. 

If $\beta_0, \beta >0$ are tuned optimal to minimize $L_{f_1}, L_{f_2}$ independently based on an off-the-shelf curve fitting solver, then we have $L_{f_2} << L_{f_1}$. (the optimal LOGIT beats the optimal linear function in all seen cases by orders of magnitude). 
\end{myfact}
\begin{myproof}
This fact is validated in Appendix \S\ref{proof:fact}, where we present Tables~\ref{tab:linear-params}--\ref{tab:logit-params} with the optimal $\beta_0, \beta >0$ for $f_1, f_2$ over each $\bar{\omega}_2$ and the corresponding $L_{f_1}, L_{f_2}$. $\blacksquare$
\end{myproof}

\begin{figure}[t]\centering
\begin{tikzpicture}[font=\sffamily,scale=0.95]
\begin{axis}
[
xmajorgrids = true,
ymajorgrids = true,
height=5.3cm,
width=6.35cm,
x label style = { yshift=0.01cm },
y label style = { yshift=-0.20cm },
xlabel= \textsf{entropy of evidence $H$},ylabel= \textsf{weight of evidence $\omega_2(H)$},xtick={0,0.1,...,1.1},ytick={0,0.1,...,1.1},
xmin=0,xmax=1.0,ymin=0,ymax=1,tick,samples=500,
domain=0:1.1,restrict y to domain =0:1.15,scale=1.4,legend style={at={(0.56,0.42)}}
]
  \addplot[ultra thick,domain=0:1,loosely dotted, mark size=1.0pt,variable=\x,green] plot ({\x},{1});  
  \addplot[very thick,domain=0:0.75,densely dotted, mark size=1.0pt,variable=\x,blue!70] plot ({\x},{1});  
 \addplot[thick,domain=0:1.6,dotted, mark size=1.0pt,variable=\x,violet!100] plot ({\x},{1-\x});  
 \addplot[thick,domain=0:1,solid,variable=\x,black] plot ({\x},{1/(1+exp(-50.27+68.766*\x))});
  \addplot[thick,domain=0:1,dashdotted, mark size=1.0pt,variable=\x,red!90] plot ({\x},{1/(1+exp(-2.75+5.5*\x))});
  \addplot[ultra thick,domain=0.75:1,densely dotted, variable=\x,blue!70] plot ({\x},{0});  
\addplot[style=very thick, mark=*, mark size=1.0pt,color=blue!70]   table [x=h, y=w, col sep=comma] {\stepf};
\addplot +[color=black!100, mark=none, dashed] coordinates {(0.75, 0) (0.75, 1)};
\addplot +[color=black!100, mark=none, dashed] coordinates {(0, 0.26) (1, 0.26)};
\legend{\footnotesize{MLE},\footnotesize{STEP$_{D=0.75}$}, \footnotesize{LINEAR}, \footnotesize{LOGIT$_3(\vec{x}_1$)}, \footnotesize{LOGIT (approx.$\!\!$ linear)}}
\end{axis}
\end{tikzpicture}
\vspace{-5pt}
\caption{Various weighting profiles from MLE and STEP to LINEAR. Profile LOGIT$_3(\vec{x}_1
)$, with $|P|\!=\!3$, is fit to prior $\vec{x}_1$ 
(Table~\ref{tab:logit-params}) 
with $\beta_0\!=\!50.270,\, \beta\!=\!68.766$. LOGIT profile (approx.$\!\!$ linear) is arbitrary $\!$with, say, $\beta_0\!=\!2.75, \beta\!=\!5.5$, to show LOGIT's flexibility. 
}
\label{fig:profiles}
\end{figure}
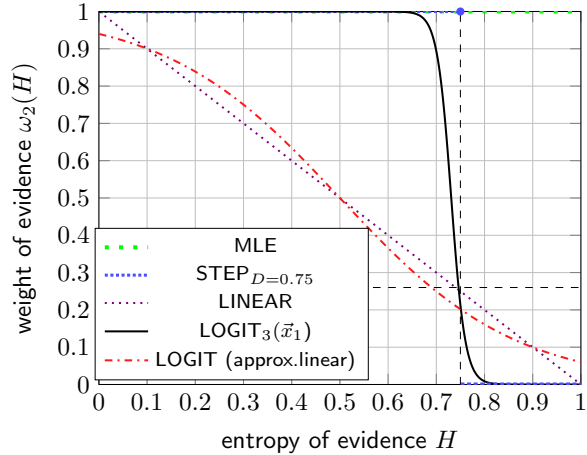

Fact~\ref{fact:logit} has been verified with several other values $0 < D < 1$, and is decisive about which weighting profile to choose. See Fig.~\ref{fig:profiles} for a visual reference, and note that it shows all the weighting profiles we have seen so far. 

We observe that LINEAR, in particular, could only have a small $L_f$ given some $(D, \bar{\omega}_2)$ if this point happens to be close enough to the line $\omega_2(H)=1-H$. 
For this reason, entry $\vec{x}_2$ for $(.75, .260)$ in Table~\ref{tab:linear-params} has the smallest $L_f$. 
Even if we seek another linear profile with $\beta_0 \neq \beta \neq 1$, we would have to deviate from the other constraints $(0, 1), (1, 0)$. LOGIT, on the other hand, is also up to 2 parameters and, most importantly, it can fit the third point anywhere in the plane with smaller deviation from $(0, 1), (1, 0)$. LOGIT can have even a nearly linear shape if so desired.

Even after going through the specified arbitrary point, LOGIT still has one remaining degree of freedom, which determines how fast it changes.  At one extreme, it could be almost linear.  At another extreme, it could be almost a step function.  Ideally, we want it somewhere in between.  We manage this by setting the slope of the function near the ambiguity threshold.  Intuitively, we are choosing the \emph{slope} near the ambiguity threshold in addition to the \emph{intercept information} $(D, \bar{\omega}_2)$.
For this purpose, we define a fourth constraint $(D + \delta, \bar{\omega}_2^\prime)$, where $\delta \!\in \mathbb R$ is small (e.g., $\delta=\!.05$) and defines a neighborhood of $D^+$. We let $P^\prime := P \cup \{(D\!+\!\delta,\, \bar{\omega}_2^\prime)\}$. This value $\bar{\omega}_2^\prime$ is taken from the inverse slope $-m^{-1}$, where slope $m$ is positive and computed as usual $m=|\bar{\omega}_2(D+\delta)\!-\!\bar{\omega}_2(D)|\,/\,(D\!+\!\delta\!-\!D)$. The intuition, as just mentioned, is to enforce the opposite behavior w.r.t. Fig.~\ref{fig:priors}: the more vertical the slope is, the smoother the weighting profile can be. Fact~\ref{fact:logit} is invariant to the extension of $P$ to $P^\prime$. 

The above reasoning is encoded into Algorithm~\ref{alg:profiling}. 
In Appendix~\ref{subsec:tuned-logit} we give final tuned parameters for LOGIT$_4$, 
with $|P|\!=\!4$ where $P^\prime\!=\!\{(0, 1),\, (1, 0),\, (.75, \bar{\omega}_2), (.80, \bar{\omega}_2^\prime)\}$. The ambiguity threshold value of $D=.75$ is discussed later in \S\ref{sec:eval}.

\begin{myrmk}$\!\!\!$
We propose \emph{\textbf{LOGIT}} to be built into \textsf{Bsmooth} (see next). Overall, we have met inverse correlation (cf. Remark~\ref{rmk:linear}) and been able to extract information from the DB-intrinsic scoring in spite of its limited informativeness (cf. Table~\ref{tab:entropy} in \S\ref{sec:entropy}). We will revisit these intuitions in \S\ref{subsec:accuracy}. $\Box$
\end{myrmk}

\vspace{-15pt}
\begin{algorithm}[b]\footnotesize
\caption{Prior's profiling. }
\label{alg:profiling}
\begin{algorithmic}[1]
\Procedure{PriorsProfiling}{$\vec{x}\!\!:\, \text{\small{scoring vector}};\; D\!\!:\, \text{ambiguity thres.};\;\delta\!\!:\, \text{neighbd.};\; \mathcal S\!\!:\, \text{population of terms}$}
\vspace{-2pt}
\Statex{}
\hrulefill
\Statex{\textbf{Part I: weighting threshold intercept}}
\State find entropy mean $\bar{H}(\vec{x})$ over $\mathcal S\!\!$  \Comment{applies Eq.~\ref{eq:shannon}}
\State find $\vec{\theta}_{prior}(x_1,\, x_2)$ such that $H(\theta_{prior}) \approx \bar{H}(\vec{x})$
\State find $\vec{\theta}_{evid}(y_1,\, y_2)$ such that $H(\theta_{evid}) \approx D$\vspace{2pt}
\State $\bar{\omega}_2 \gets \lambda/(1+\lambda)$, where $\lambda=|x_1-x_2|\,/\,|y_1-y_2|$\vspace{2pt}
\State $P \gets \{(0, 1),\, (1, 0),\, (D, \bar{\omega}_2)\}$ \Comment{initializes $P$}
\vspace{-2pt}
\Statex{}
\hrulefill
\Statex{\textbf{Part II: weighting threshold slope}}\vspace{2pt}
\State $m \gets |\,\bar{\omega}_2(D+\delta)-\bar{\omega}_2(D)\,|\,/\,\delta$\Comment{$\delta\!=\!(\cancel{D}\!+\!\delta)\!-\!\cancel{D}$}\vspace{1pt}
\State $\bar{\omega}_2^\prime \gets \bar{\omega}_2(D) -m^{-1}\,\delta$\Comment{intercept $\!-\!$ slope$^{-1}$ $\!\cdot\; \delta$}\vspace{1pt}
\State $P^\prime \gets P \cup \{(D+\delta,\, \bar{\omega}_2^\prime)\}$\vspace{2pt}
\State find $\beta_0,\, \beta>0\;$ such that LOGIT's lack of fit to $P^\prime$ is min.
\State \Return $\beta_0,\, \beta$
\EndProcedure
\end{algorithmic}
\end{algorithm}
\vspace{-5pt}

\vspace{25pt}
\subsection{The Bayesian Smoothing Algorithm}\label{subsec:smoothing}
\noindent
The complete pipeline is encoded into Algorithms~\ref{alg:profiling}-\ref{alg:smoothing}. A high level description of our approach is as follows. 

\begin{enumerate}
\item Weight of evidence $\omega_2$ is really a predictive probability, $p( \vec{y})$, about whether the extrinsic source $\vec{y}$ is trustable. To do bayesian smoothing wisely, we need to find $\omega_2(H)$, i.e., to find $p(\, \vec{y} \,|\, H(\vec{y}) \,)$, which should have a LOGIT profile.

\item Besides the two obvious points $\{(0, 1),\, (1, 0)\}$ in the plane $[0, 1]^2$, we need to fit one or two more constraints to define the behavior of our weighting scheme at the ambiguity threshold, $H_D$, and nearby, $H_D+\delta$. 

\item Intuitively, since $H_D$ is the turning point of regularity for (in-)accuracy, at $H_D$ the weighting scheme should combine our two sources towards balance. Theorem~\ref{thm:convex} gives theoretical support by defining conditions for balance.

\item Now given one's own specific prior profile we can find constraints $P^\prime\!=\!\{(0, 1),\, (1, 0),\, (H_D,\, \bar{\omega}_2),$ $(H_D+\delta,\,\, \bar{\omega}_2^\prime)\}$. This is done (Algorithm~\ref{alg:profiling}), only once, offline. The fitted logistic curve is just $p(\, \vec{y} \,|\, H(\vec{y}) )$, which we were seeking. It provides parameters $\beta_0,\, \beta >0$ into Algorithm~\ref{alg:smoothing}.

\item Given a new term disambiguation problem instance, by Algorithm~\ref{alg:smoothing} we simply compute $H(\vec{y})$ and get $p( \vec{y} \,|\, H(\vec{y}) )$, namely, $\omega_2$, and then apply bayesian smoothing given $\omega_2$. 
\end{enumerate}

\begin{algorithm}[t]\footnotesize
\caption{Bayesian smoothing. }
\label{alg:smoothing}
\begin{algorithmic}[1]
\Procedure{Bsmooth}{$\vec{x},\, \vec{n}\!\!:\, \text{\small{vectors}};\;\, \beta_0, \beta\!\!:\, \mathbb R_{>0}$}
\Ensure returns a convex combination $\vec{\theta}= \vec{x} \oplus \vec{y}$
\State $\vec{y} \gets \vec{n}\;$ scaled to 1 
\State $H \gets \left( \sum_i^k{ -y_i\,\log{y_i} } \right)/\log{k}$ \Comment{Shannon's Eq.~\ref{eq:shannon}}
\State $\omega_2 \gets 1\,/\,(1+\exp{(-\beta_0 + \beta\,H)}\,)$\Comment{LOGIT (Eq.~\ref{eq:logit})}
\State $\omega_1 \gets 1- \omega_2 $ 
\State $\alpha \gets n\, (\omega_1/\omega_2)$ \Comment{sets the scale of $\alpha$}
\State $\vec{\alpha} \gets \vec{x}\;$ scaled to $\alpha$ 
\For{$i \gets 1 \text{\;to\;} k$}  \Comment{Dirichlet-multinomial (Eq.~\ref{eq:dm})}
\State $\theta_i \gets (n_i + \alpha_i)/(n+\alpha)$
\EndFor
\State $\vec{\theta} \gets \vec{\theta}\;$ scaled to 1 \vspace{1pt}
\State \Return $\vec{\theta}$
\EndProcedure
\end{algorithmic}
\end{algorithm}

Clearly, \textsf{Bsmooth} is bounded in time $O(k)$, where $k$ is the number types retrieved, fixed in query and data complexity. Considering IMDb and other benchmarked databases \cite{coffman2010}, a typical value is $k=5$, and yet for a very complex DB schema one could think of $k \leq 5\,c$ for some small $c \in \mathbb N^\ast$.

\vspace{-2pt}
\section{Effectiveness Evaluation}\label{sec:eval}
\noindent
We use the benchmark dataset D5-BENCH (cf. \S\ref{sec:settings}), developed independently \cite{coffman2010,coffman2014}, as a gold standard to evaluate accuracy rates in several settings.

\subsection{Evaluation Metrics}\label{subsec:metrics}
\noindent
The accuracy metric that is central to this paper is P@1 \cite{manning2008}. This is because the real system that we are modeling must avoid mistakes on the default option that is presented to users as disambiguation to a query term. Note that every accuracy mistake burdens the user with an interaction that is then necessary to satisfy her information need. `Recall,' by comparison, is not important because $k \in \mathbb N$, the number of candidate schema elements that are potentially relevant, is bound to be small.
To make sure P@1 is not too rigorous, we will also cover (the more flexible) `mean reciprocal rank.' 

\textbf{Precision-at-rank--$1$ (P@1)}. For each term P@1 scores 1 when the highest-ranked result is relevant, and 0 otherwise. For the ranking shown for query term `tom hanks' in Example~\ref{ex:hanks}, we have P@1 zero.
We will aggregate P@1 by arithmetic mean over the 62 terms. 

\textbf{Mean Reciprocal Rank (MRR)}. This is the reciprocal of the highest-ranked relevant result for each query term averaged over all of them (see Eq. \ref{eq:mrr}). In the ranking shown for `tom hanks' in the first part of Example~\ref{ex:hanks}, we have RR=1/2 because the one relevant result \textsf{NAME.name} is ranked at position 2. 
\begin{equation}\label{eq:mrr}
MRR = \frac{1}{n} \, \displaystyle\sum_{i=1}^{n} \frac{1}{rank(i)}
\end{equation}

\subsection{Study of Accuracy Profiles}\label{subsec:accuracy}
\noindent
We have designed \textsf{Bsmooth} under assumption that entropy is informative for the extrinsic source (evidence) but not for the intrinsic one (hypothesis). Now it is time to revisit and see if it carries on for interactive data retrieval.

Fig.~\ref{fig:histograms} shows P@1 scores for D1-INTR and D3-IMPL across ranges of increasing entropy. Given the entropy of the scoring vector for each of the 62 terms, we formed 5 entropy ranges of sizes 12, 12, 12, 13, 13, covering D1 and D3 from the lowest- to the highest-entropy scored terms. 
We observe: 

\begin{itemize}
\item For D1-INTR (37/62 hits in P@1), no (inverse) correlation is seen between entropy and accuracy. The intrinsic model is subject to noise in unpredictable ways. 

\item For D3-IMPL (54/62 hits in P@1), we do observe inverse correlation, which does not seem to be linear (6 mishits in the last range and 1 in the range before). Note carefully that it is only from some entropy threshold on that we observe a significant drop in accuracy. 
\end{itemize}


We see that implicit user feedback (D3) have high P@1 accuracy. So the user interactions really qualify for `observed' counts for bayesian inference --- this is a non-obvious finding for interactive data retrieval.


\begin{figure}[t]\scriptsize
\vspace{-1pt}
\advance\leftskip 2.5cm
\begin{tikzpicture}[font=\small,scale=0.85]
    \begin{axis}[
      width=10cm,
      height=3.5cm,
      ybar,
      ylabel={\large{P@1}},
      xlabel={\large{entropy range}},
      y label style = { yshift=-0.4cm },
      x label style = { yshift=0.05cm },
      ytick={0,2,...,12},
      ymajorgrids = true,
      ymin=0,
      ytick=\empty,
      xtick=data,
      axis x line=bottom, 
      axis y line=left,
      enlarge x limits=0.3,
      bar width=8pt,
      nodes near coords={\pgfmathprintnumber\pgfplotspointmeta},
      legend style={font=\small,at={(1.12,1.17)}}
]
\legend{D1-INTR, D3-IMPL}
\addplot [] coordinates {(1,5) (2,7) (3,10) (4,9) (5,6)};
\addplot [pattern=horizontal lines dark gray] coordinates {(1,12) (2,12) (3,12) (4,12) (5,6)};
\end{axis}
\end{tikzpicture}
\caption{Histograms of P@1 hits over ranges of increasing entropy for datasets D1-INTR and D3-IMPL.}
\label{fig:histograms}
\end{figure}
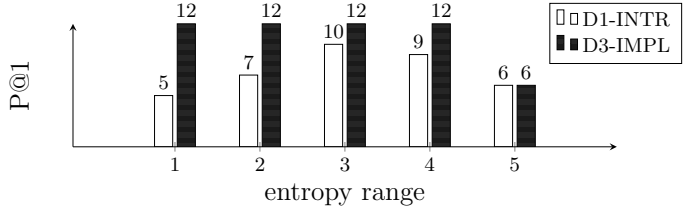

\begin{figure*}[t]
\advance\leftskip-0.35cm
\begin{tikzpicture}[scale=0.9]
  \begin{axis}[
    bar width=7pt,
    width=18.0cm,
    height=4.25cm,
    xtick=data,
    ymin=0,
    ymax=100,
    ybar=3pt,
    ytick={0,25,...,100},
    ymajorgrids = true,
    x axis line style = { opacity = 0 },
    xticklabel style={anchor=base,yshift=-0.15cm},
    axis y line*=left,
    tickwidth = 0pt,
    enlarge x limits  = 0.25,
    enlarge y limits  = 0.1,
    symbolic x coords = {P@1.EXPL, P@1.IMPL, P@1.RAND, MRR.EXPL, MRR.IMPL, MRR.RAND},
    xticklabel style={anchor=base,rotate=0,yshift=-0.10cm},
    every node near coord/.append style={font=\small},
   nodes near coords,
    legend style={font=\small,at={(1.03,1.05)}}
  ]
\addplot [] coordinates { (P@1.EXPL, 60) (MRR.EXPL, 78) (P@1.IMPL, 60) (MRR.IMPL, 78) (P@1.RAND, 60) (MRR.RAND, 78) }; 
  \addplot [pattern=horizontal lines] coordinates { (P@1.EXPL, 97) (MRR.EXPL, 98) (P@1.IMPL, 87) (MRR.IMPL, 92) (P@1.RAND, 13) (MRR.RAND, 39) };
  \addplot [pattern=dots] coordinates { (P@1.EXPL, 92) (MRR.EXPL, 96) (P@1.IMPL, 82) (MRR.IMPL, 90)  (P@1.RAND, 48) (MRR.RAND, 70) }; 
  \addplot [pattern=north west lines] coordinates { (P@1.EXPL, 97) (MRR.EXPL, 98) (P@1.IMPL, 89) (MRR.IMPL, 94) (P@1.RAND, 39) (MRR.RAND, 64) }; 
  \addplot [pattern=horizontal lines dark gray] coordinates { (P@1.EXPL, 95) (MRR.EXPL, 98) (P@1.IMPL, 90) (MRR.IMPL, 95) (P@1.RAND, 58) (MRR.RAND, 75) };
    \legend{INTR, MLE, STEP, LINEAR, LOGIT}
  \end{axis}
\end{tikzpicture}
\caption{P@1 and MRR in \% over the $n\!=\!62$ query terms with number of retrieved types $k\!=\!5$. $\!$The metrics are shown for baseline choices INTR, MLE, STEP, LINEAR and proposed technique LOGIT over datasets D2-EXPL, $\!$D3-IMPL, $\!$D4-RAND.}
\label{fig:internal-eval}
\end{figure*}
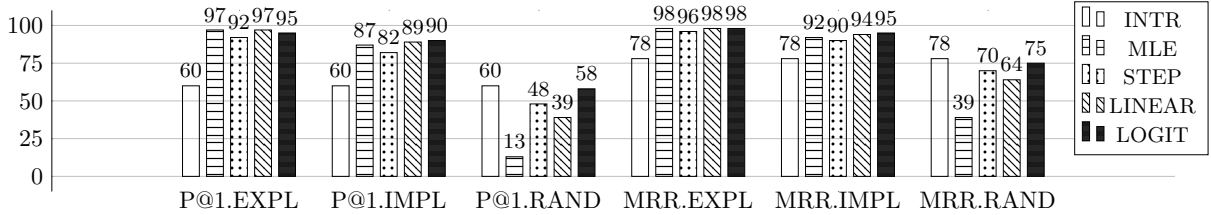

\subsection{Internal Evaluation}\label{subsec:internal}
\noindent
The histogram for D3-IMPL in Fig.~\ref{fig:histograms} provides empirical support to set the ambiguity threshold $H_D\!\approx\!.75$. The results we present here are based on this tuning.
\footnote{To review and fine tune their own ambiguity threshold, users should simply plot an histogram like Fig.'s~\ref{fig:histograms} for the variant of D3-IMPL obtained from their own interaction log, either real or simulated, given the golden answer keys from their own variant of D5-BENCH. } 

Fig.~\ref{fig:internal-eval} shows results of the internal evaluation. We concentrate the discussion on P@1 (left), as MRR (right) is shown only to make sure the P@1 results preserve their structure in a more flexible metric --- and they do. 

The main observation is that LOGIT, the proposed bayesian smoothing strategy, can roughly retain the accuracy of INTR even under D4-RAND data, while significantly beat it under both D2-EXPL and D3-IMPL data. Its P@1 scores are also better than the baseline strategies MLE, STEP and LINEAR in the most important datasets D4-RAND and D3-IMPL, which simulate user feedback with skepticism and avoid user burden. The differences in P@1, of course, are not large because we have deliberately moved from one strategy to the other in seek of every small improvement.

STEP lacks a continuous decay profile at the neighborhood of the ambiguity threshold, and LINEAR fails to penalize high entropy fast enough. MLE is in fact not solid to be deployed into a system strategy. We see that under RAND data, say when a term is very ambiguous or users behave undecidedly as a crowd, it would make the system vulnerable. INTR is more often accurate than not (60\%), yet not highly accurate. It can be significantly improved by learning from implicit feedback as an extrinsic source. 
These are robust results towards LOGIT-based bayesian smoothing.

\begin{figure}[t]
\vspace{-6pt}
\advance\leftskip 2.0cm
\begin{tikzpicture}[scale=0.8]
  \begin{axis}[
    bar width=2.5mm,
    width=12cm,
    height=7cm,
    xtick=data,
    ybar,
    ymin=0,
   xmin=P@1,
   xmax=MRR,
    ymax=100,
    ytick={0,25,...,100},
    ymajorgrids = true,
     x axis line style = { opacity = 0 },
    axis y line*=right,
    tickwidth = 0pt,
    enlarge x limits  = 0.58,
    enlarge y limits  = 0.1,
    symbolic x coords = {P@1, MRR},
    xticklabel style={anchor=base,yshift=-0.10cm},
    every node near coord/.append style={font=\small},
   nodes near coords,
    legend style={font=\small,at={(0.3,1.36)}}
  ]
  \addplot [pattern=grid] coordinates { (P@1, 17) (MRR, 19) };
  \addplot [pattern=crosshatch] coordinates { (P@1, 12) (MRR, 16) };
  \addplot [pattern=fivepointed stars] coordinates { (P@1, 0) (MRR, 0) };
  \addplot [pattern=vertical lines] coordinates { (P@1, 08) (MRR, 15) };
  \addplot [pattern=horizontal lines] coordinates { (P@1, 08) (MRR, 09) }; 
  \addplot [pattern=north east lines] coordinates { (P@1, 06) (MRR, 14) }; 
  \addplot [pattern=sixpointed stars] coordinates { (P@1, 0) (MRR, 0) }; 
  \addplot [pattern=dots] coordinates { (P@1, 10) (MRR, 15) }; 
  \addplot [pattern=sixpointed stars] coordinates { (P@1, 0) (MRR, 0) }; 
  \addplot [pattern=north west lines] coordinates { (P@1, 28) (MRR, 34) }; 
  \addplot [] coordinates { (P@1, 60) (MRR, 77) }; 
  \addplot [fill=gray] coordinates { (P@1, 95) (MRR, 98) }; 
  \addplot [fill=black] coordinates { (P@1, 90) (MRR, 95) }; 
    \legend{BANKS \cite{banks2002}, DISCOVER \cite{discover2002}, BANKS-II \cite{banksii2005}, DISCOVER-II \cite{discoverii2003}, Liu et al. \cite{liu2006}, DPBF \cite{dbpf2007}, BLINKS \cite{blinks2007}, SPARK \cite{spark2007}, STAR \cite{star2009}, CD \cite{coffman2010b}, INTR \cite{li2015}, LOGIT/EXPL, LOGIT/IMPL}
  \end{axis}
\end{tikzpicture}
\caption{P@1 and MRR over 50 queries with 100 retrieved tuples each. The metrics are shown for related systems in the literature (source: \protect\cite{coffman2014}) and LOGIT/EXPL and /IMPL.}
\label{fig:external-eval}
\end{figure}
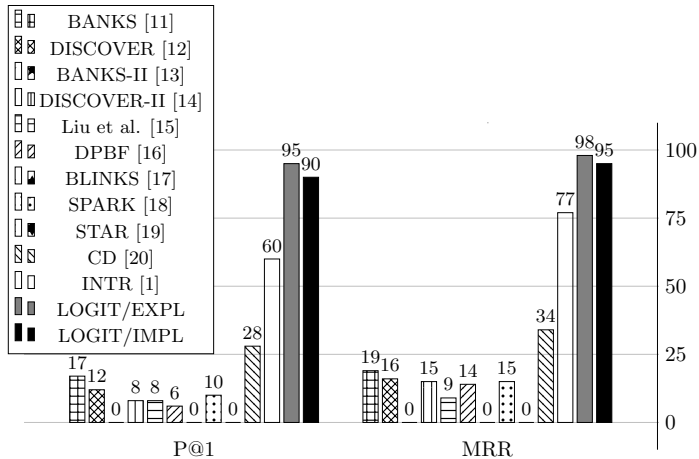

\subsection{External Evaluation}\label{subsec:external}
\noindent
We have also obtained from the benchmark's authors the accuracy rates of 10 competitive systems in the literature on the exact same IMDb dataset (D0) and 50 queries \cite{coffman2014}. We do not have, however, the specific tuples returned in order to `aggregate' them into types. The `external' evaluation comparison is then indirect in the following sense.

To project our accuracy rates onto tuple level, we assume an accuracy hit in P@1 only if we disambiguate accurately all the terms in a query. Otherwise we take P@1 zero and RR to be the least over all terms in that query. 
Fig.~\ref{fig:external-eval} shows the accuracy rates for the external evaluation. Not surprisingly, the addition of a learning layer (LOGIT/D3-IMPL) on top of a DB-intrinsic source (INTR) strongly outperforms the literature systems \cite{banks2002,discover2002,banksii2005,discoverii2003,liu2006,dbpf2007,blinks2007,spark2007,star2009,coffman2010b}.

 When a system has score zero, it is because according to the benchmark authors (not us) it takes a really large time to respond \cite{coffman2014}. 
Unpredictable response times are also reported by Baid et al. for many of these systems \cite{baid2010}. Although we are not concerned with performance evaluation here, it is a fair point that unacceptable response time is prohibitive for effectiveness \cite{coffman2014}. Excessive response time is pointed out as related with the generation of large data graphs \cite{baid2010}. 

An in-depth evaluation of those systems is presented in \cite{coffman2014}. Here we are concerned with evaluating a learning layer that is fairly independent of any particular method to produce the DB-intrinsic scoring, and takes only $O(k)$ time  (cf. \S\ref{subsec:smoothing}).

\section{Empirical Analysis of Implicit Feedback}\label{sec:implicit}
\noindent
Recall that explicit user feedback is desirable, but may be unavailable or limited.  As such, we expect to be limited to implicit user feedback in most practical situations.
We saw above that \textsf{Bsmooth}'s accuracy with D3-IMPL was only slightly worse than its accuracy with D2-EXPL.
In this section, we look at this issue more carefully.

The central problem with implicit feedback is to interpret correctly the case where a worker just accepts the default choice, as if in session abandonment. 
How could we know whether this happened due to user satisfaction with the default or user indifference (or `laziness')? 
Fig. \ref{fig:probability-tree} shows the corresponding estimation scheme as a probability tree. There are two parameters: $\alpha:= P(A)$ and $\ell:= P(L)$, where $A$ is the event of abandonment and $L$ the event of `laziness.'

\tikzstyle{level 1}=[level distance=3.6cm, sibling distance=3.5cm]
\tikzstyle{level 2}=[level distance=3.5cm, sibling distance=3.0cm]
\tikzstyle{bag} = [text width=4em, text centered]
\tikzstyle{end} = [circle, minimum width=3pt,fill, inner sep=0pt]
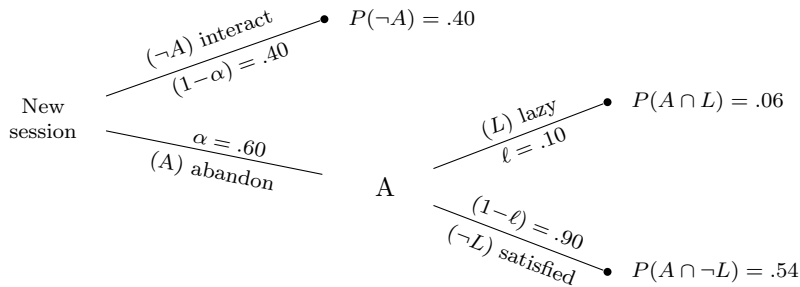
\begin{figure}[t]
\advance\leftskip 1.5cm
\begin{tikzpicture}[grow=right, sloped,scale=0.75]
\node[bag,font=\footnotesize] {New session}
    child {
        node[bag, xshift=1.8cm, yshift=0.4cm] {A}        
            child {
                node[xshift=0.3cm, end, label=right:
                    {\footnotesize{$\;\,P(A \cap \neg L)=.54$}}] {}
                edge from parent
                node[above,yshift=-0.05cm, font=\footnotesize] {$\;(1\!-\!\ell)=.90$}
                node[below,font=\footnotesize]  {$(\neg L)$ satisfied}
            }
            child {
                node[xshift=0.3cm, end,font=\footnotesize, label=right:
                    {\footnotesize{$\;\,P(A \cap L)=.06$}}] {}
                edge from parent
                node[xshift=0.1cm, yshift=-0.05cm, above,font=\footnotesize] {$(L)$ lazy}
                node[xshift=0.1cm, yshift=0.05cm, below,font=\footnotesize]  {$\;\ell=.10$}
            }
            edge from parent
            node[xshift=0.15cm, yshift=-0.05cm, font=\footnotesize, above] {$\,\alpha=.60$}
            node[xshift=0.0cm, below,font=\footnotesize]  {$\,$(A) abandon}
    }
    child {
        node[xshift=1cm, end, font=\footnotesize, label=right: {\footnotesize{$\;\,P(\neg A)=.40$}}] {}
        edge from parent         
            node[above, xshift=0.0cm, font=\footnotesize] {$\;(\neg A)$ interact}
            node[below, yshift=0.05cm, font=\footnotesize, xshift=0.1cm]  {$\;\,(1\!-\!\alpha)=.40$}
    };
\end{tikzpicture}
\caption{Probability tree diagram and a valuation example. The abandonment rate ($\alpha$) and `good' abandonment rate $\alpha\,(1\!-\!\ell$) are free parameters.} 
\label{fig:probability-tree}
\end{figure}

We can observe the rate of session abandonment, $\alpha$, but not the `laziness' parameter $\ell$. 
If we have some explicit feedback available, that can be used as a baseline to estimate $\ell$.  
We need to see if there is a significant `laziness' effect behind implicit feedback. So for each term, e.g., `angelina jolie', we take the default option (\textsf{NAME.name}) shown to the worker for the generation of D3-IMPL, and compared the feedback rate (with possible laziness effect) of that particular option against its corresponding rate from D2-EXPL (baseline). 

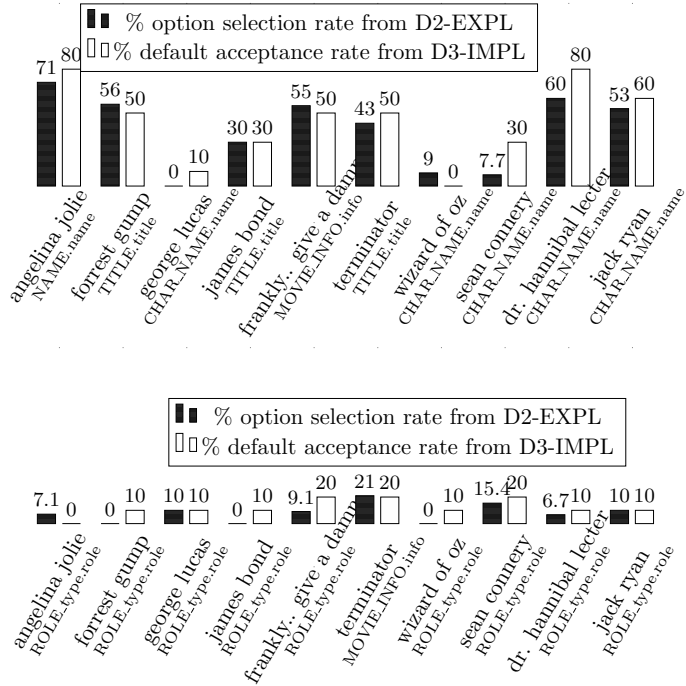
\begin{figure}[t]
\advance\leftskip 2.0cm
\begin{subfigure}{.99\textwidth}
\begin{tikzpicture}[scale=0.85]
  \begin{axis}[
xtick={1, 2, 3, 4, 5, 6, 7, 8, 9, 10},
xticklabels={\textunderset{NAME.name}{angelina jolie}, \textunderset{TITLE.title}{forrest gump}, \textunderset{CHAR\_NAME.name}{george lucas}, \textunderset{TITLE.title}{james bond}, \textunderset{MOVIE\_INFO.info}{frankly.. give a damn}, \textunderset{TITLE.title}{terminator}, \textunderset{CHAR\_NAME.name}{wizard of oz}, \textunderset{CHAR\_NAME.name}{sean connery}, \textunderset{CHAR\_NAME.name}{dr. hannibal lecter}, \textunderset{CHAR\_NAME.name}{jack ryan}
},
    bar width=8pt,
    width=14cm,
    height=5cm,
    xtick=data,
    ymin=0,
    ymax=100,
    ybar=3pt,
    x axis line style = { opacity = 0 },
    axis y line       = none,
    tickwidth         = 0pt,
    enlarge x limits  = 0.2,
    enlarge y limits  = 0.25,
      xticklabel style={anchor=base,rotate=57,yshift=-0.15cm,xshift=-0.4cm},
    every node near coord/.append style={font=\small},
   nodes near coords,
    legend style={at={(0.74,1)}} ]
\addplot [pattern=horizontal lines dark gray] coordinates { (1,71) (2,56) (3,0) (4,30) (5,55) (6,43) (7,9) (8,7.7) (9, 60) (10,53) };
  \addplot [] coordinates { (1,80) (2,50) (3,10) (4,30) (5,50) (6,50) (7,0) (8,30) (9, 80) (10, 60) };
    \legend{\% option selection rate from D2-EXPL, \% default acceptance rate from D3-IMPL}
  \end{axis}
\end{tikzpicture}
\end{subfigure}
\begin{subfigure}{.99\textwidth}
\vspace{-5pt}
\begin{tikzpicture}[scale=0.85]
  \begin{axis}[
xtick={1, 2, 3, 4, 5, 6, 7, 8, 9, 10},
xticklabels={\textunderset{ROLE\_type.role}{angelina jolie}, \textunderset{ROLE\_type.role}{forrest gump}, \textunderset{ROLE\_type.role}{george lucas}, \textunderset{ROLE\_type.role}{james bond}, \textunderset{ROLE\_type.role}{frankly.. give a damn}, \textunderset{MOVIE\_INFO.info}{terminator}, \textunderset{ROLE\_type.role}{wizard of oz}, \textunderset{ROLE\_type.role}{sean connery}, \textunderset{ROLE\_type.role}{dr. hannibal lecter}, \textunderset{ROLE\_type.role}{jack ryan}
},
    bar width=8pt,
    width=14cm,
    height=5cm,
    xtick=data,
    ybar=3pt,
    ymin=0,
    ymax=100,
    x axis line style = { opacity = 0 },
    axis y line       = none,
    tickwidth         = 0pt,
    enlarge x limits  = 0.2,
    enlarge y limits  = 0.325,
      xticklabel style={anchor=base,rotate=60,yshift=-0.15cm,xshift=-0.4cm},
    every node near coord/.append style={font=\small},
   nodes near coords,
    legend style={at={(0.85,0.77)}} ]
\addplot [pattern=horizontal lines dark gray] coordinates { (1,7.1) (2,0) (3,10) (4,0) (5,9.1) (6,21) (7,0) (8,15.4) (9, 6.7) (10,10) };
  \addplot[] coordinates { (1,0) (2,10) (3,10) (4,10) (5,20) (6,20) (7,10) (8,20) (9, 10) (10,10) };
    \legend{\% option selection rate from D2-EXPL, \% default acceptance rate from D3-IMPL}
  \end{axis}
\end{tikzpicture}
\end{subfigure}
\vspace{-12pt}
\caption{Explicit and implicit choice selection rates, with a good default (top row) and bad default  (bottom row). Even though there is considerable variability across terms, there is a clear lazy abandonment gap of about 5\%.}
\label{fig:rates}
\vspace{-12pt}
\end{figure}

Fig. \ref{fig:rates} (top) shows the results, term-wise, for a sample of 10 terms. For query term `angelina jolie,' we see a D2-score of 71\%, and a D3-score of 80\%. Thus for this particular term we have an estimated laziness of $\ell= 9\%$. Across the other 9 terms, we see, of course, some variance on $\ell$. Over all the 62 terms, we have mean $\ell=5.1 \pm 13\%$. 

Now, to make sure the laziness effect carries on even when a different default option is used, we ran an additional experiment with the implicit feedback set up. 
Since D3 has been produced by presenting as default option the highest scored option from D1-INTR, such additional check would be most effective to stress $\ell$ if we now show D1's lowest scored option as default. We rendered a dataset D3$^\prime$ this way, as a variant of D3 where we knew the default answer was wrong.  

Specifically, we instantiated a task to 10 new workers for each query term, and showed them such a `bad' default. 
Fig. \ref{fig:rates} (bottom) shows results based on D3$^\prime$. 
We also see fairly close explicit and implicit feedback rates. For term `forrest gump,' e.g., the `bad' default was \textsf{ROLE\_TYPE.role}, which obtained D2-score 0\% and D3$^\prime$-score 10\%. For this term we have an estimated laziness of $\ell= 10\%$.   
Over all the 62 terms, we have mean $\ell=4.8 \pm 8\%$. 

In sum, first, in both cases the rates from explicit (D2) and implicit feedback (D3 or D3$^\prime$) are similar termwise. Second, the average laziness rates (resp., 5.1\%, and 4.8\%) are compatible across the two different settings. 
That is, if our crowd tasks have a reasonable design to simulate data retrieval interactions then the laziness parameter has empirical support to be set $\ell \approx 5\%$. On average, therefore, users have only a 5\% bias towards just accepting the default choice whatever it is.  
In other words, implicit feedback is expected to work nearly as well as explicit feedback, which is exactly what we observed in the evaluation of \S\ref{sec:eval}. 

\vspace{-4pt}
\section{Related Work}\label{sec:related-work}
\noindent
\textbf{Machine learning models}. Problems like ours are often addressed by learning a classifier from labeled training data, and then applying it to the unseen instances.\footnote{Were the disambiguation response variable real-valued and not categorical then the problem would be one of  regression, not classification \cite{murphy2012}.} This is a popular approach, when (i) the training data is really large and diverse enough to avoid overfitting; and (ii) no insight is available other than the training data. Recall, however, that (i) D5 has only 62 examples (labeled queried terms).
Having such low number is common since acquiring labeled data is expensive; and the interaction log for the queried terms population may form a `long tail,' possibly making the real system log small termwise; also, (ii) D1-INTR is available with 60\% P@1 accuracy; it is therefore a clearly useful information source to build upon. 

So the IDR problem setting and how to make the best of its two sources is more challenging and does not suit standard supervised learning. That is why we have taken a more original approach for term classification. 

\textbf{Combination of experts}. 
Many models combine forecasts from several human experts as independent information sources on uncertain events. 
They give their probability estimates, which would be compatible with relative frequencies obtained from multinomial counts like in a user interaction log. A popular approach to combine the individual forecasts is the so-called `linear opinion pooling.' It assigns each forecast a weight that reflects the importance or quality of that expert. Recently, along these lines, more advanced techniques have been developed \cite{satopaa2014}. 
One important point of improvement is to consider `sharpness,' which rewards how close to either 0 or 1 an estimate is (ibid.), somehow in line with our choice to use entropy as a measure. 
Yet their core measure of source quality is still `calibration' (see, e.g., Brier scores).\footnote{\url{http://en.wikipedia.org/wiki/Brier_score}.} Suppose an expert estimates 0.3 as the probability for a specific outcome, then she is best calibrated if that outcome is seen in 30\% of the trials. 
Calibration is dependent on seen examples. 

Consider a scenario where the intrinsic and extrinsic sources are calibrated given D1-INTR and D3-IMPL as training data, and D5-BENCH as testing data. By Fig.~\ref{fig:histograms}, D1 turns out not to be calibrated at all, while D3 is very well calibrated. Note the structural asymmetry. 
So these ``experts'' are given (resp.) very low and very high importance. 
For term `tom hanks,' e.g., whose D1 scores and D3 log counts are shown in Example~\ref{ex:hanks}, our two ``experts'' would be given asymmetrical importance likewise. Now suppose the real system's log gets noisy because a fan of the `Bamboo Shark' movie is repeatedly querying for `tom hanks,' the name of a character in this movie, rendering the important log source effectively ``bad'' for next queries on `tom hanks,' the person's name.

Without a more complex model, e.g., to give personalized results, requiring then even more expensive training with profiled users etc, there is no way calibration could help the system to downgrade the log source and upgrade the intrinsic source ``experts'' on the fly.
Calibration can only distinguish source quality based on training data that is expensive to acquire, and yet becomes outdated: (1) as the database is updated with new records, properly captured by the `uncalibrated' intrinsic model by the way; and (2) the interaction log is updated with real use, captured by the occasionally `uncalibrated' entropy. 
Entropy is instead a unsupervised measure. It can be checked very fast every new interaction log update. For term logs that have high entropy, either due to genuine ambiguity (e.g., `terminator') or due to unusual noise, \textsf{Bsmooth} will be conservative and give more weight to the intrinsic model.


The Dempster-Shafer (DS) theory of evidence \cite{shafer1976} has been applied in related work to combine two DB-intrinsic scorings \cite{bergamaschi2013}. It does not seem to be fit to the interaction log scenario. 
The inferential scheme (Dempster's rule of combination) derives shared belief from multiple sources and ignores all the conflicting (non-shared) belief by a normalization factor. Zadeh gives an example of how counter-intuitive this is when beliefs should be rather integrated cumulatively \cite{zadeh1984}. 
Let doctors $S_1, S_2$ have beliefs for diagnosis on conditions ($A$) meningitis, ($B$) brain tumor, ($C$) concussion, viz., $S_1\!=\!\{\,P(A)\!=\!.99,\, P(B)\!=\!.01\,\}$ and $S_2\!=\!\{\,P(B)\!=\!.01,\, P(C)\!=\!.99\,\}$. Then $P(B)\!=\!1$ is inferred \cite{zadeh1984}. The doctors agree only that brain tumor is very unlikely, but such a weak consensus is pushed through anyway. Now if we think of $A, B, C$ as movies $S_1, S_2$ plan to see together then the inference finds the movie shared by their belief constraints.  
This shows that the matching between abstract framework and applied use case is really important. 

DS theory has applications of `truth finding' given conflicting and/or absent information, when sources are independent --- distributed, say, in time, space or perspective; e.g., in expert systems, question answering, data fusion, etc. The IDR use case, however, is distinguished by an asymmetry between the DB-intrinsic (model) vs. the DB-extrinsic (data) as we have extensively discussed, and seen in \S\ref{subsec:accuracy}. 

DS's belief functions to combine reliable past `data' and expert (`judgemental') evidence, e.g., as applied to sea level estimation subject to climate change \cite{denoeux2014}, or as applied to forecasts of innovation diffusion \cite{denoeux2014b}, but these differ sufficiently to the IDR use case, where no `data' is available and the `judgemental' information falls into either a certain or an ambiguous case.

\textbf{Interactive Data Retrieval}. 
As mentioned in \S\ref{sec:settings}, we used crowdsourcing as a cost-effective means to simulate user interactions and get insight to system design beforehand \cite{zuccon2013}. By two different crowd task designs, we have acquired explicit and implicit feedback and have studied them, also considering their possible limitations to feed a learning layer in IDR systems. 
Although we take some inspiration from Interactive Information Retrieval (IIR) \cite{azzopardi2010}, we avoid complex models (e.g., \cite{azzopardi2014}) to pursue and report instead a learning technique designed to be simple and very fast at query time. 
Note also that a comparison with complex models for aggregation of crowd answers is not required, since a direct measuring of its uncertainty has been enough to warrant high P@1 accuracy under a neat bayesian smoothing model tuned with LOGIT. 

Regarding implicit feedback in particular, the concept of `good abandonment' (GA) has been explored based on a large Google search log \cite{tokuda2009}. 
It indicates that GA (i.e., when the user's information need is satisfied with no need to click on a result or refine the query) may be really a significant portion of abandoned sessions. 
Considering two modalities, PC and mobile, they point out that the latter has higher GA rates. 
As mobile queries tend to be more objective, the result snippets are often enough to satisfy an information need \cite{tokuda2009}. This adds to our own findings suggesting high GA rates in the IDR use case, as a database query answer (structured list of facts) may be precise enough not to require any browsing.

\textbf{Imprecise DB queries}. 
Some ad-hoc practices in the keyword search literature that have been criticized \cite{webber2010,baid2010,coffman2014}, and that we have strived not to incur in are: 
\begin{itemize}
\item Existing scoring functions have become increasingly complex while the added value obscure \cite{coffman2014}. We have studied and shown in detail the added value of implicit feedback as a DB-extrinsic scoring to be combined by \textsf{Bsmooth} with any existing DB-intrinsic scoring.

\item Related work reports evaluation based on ad-hoc queries and databases---even arbitrary modification of schema, e.g., changing table and attribute names to match user queries better \cite{coffman2014}. We evaluate \textsf{Bsmooth} on IMDb `as is' in the benchmark with its own encoding. 

\item As mentioned, most related techniques rely on large DB-induced graphs and/or on auxiliary views for keywords, incurring in serious performance and maintenance issues. They may not apply to online databases on-the-fly \cite{baid2010}. \textsf{Bsmooth} in turn can be built at query time on top of any intrinsic model without noticeable time expense added. 

\item Most systems retrieve tuples in a one-shot response. We instead leverage on the potential of IDR with some interactions that are cheap to users. The learning layer we propose can be incorporated by any IDR system. 
\end{itemize}


\vspace{-4pt}
\section{Conclusions}\label{sec:conclusions}
\noindent
In this paper we have studied imprecise DB queries in the context of interactive data retrieval (IDR), where a little help is expected from users to improve the disambiguation of query terms. 
The potential of relying on a DB-extrinsic source in addition to the DB-intrinsic one is obvious, yet uncertainty management is not trivial. We have singled out bayesian smoothing as an adequate learning framework in the presence of noise and possibly small query logs, and seen several alternative weighting schemes that could have been built into the proposed \textsf{Bsmooth} algorithm to do it wisely. 

In fact, by simulating user interactions we have found that a crowd of users can potentially provide highly accurate feedback into a query log, even when it is collected implicitly termwise and the workers have no knowledge of the DB's internal structure. Here, one could think of storing feedback not only by term but also by query, by individual user, or even by time (sequential) leading to more complex models. Yet we have seen that none of these seem to be required to achieve highly accurate type disambiguation in IDR even given a small-size query term log. 

We have opted to advance in depth using one DB instead of in breadth with multiple DB's---the benchmark, e.g., covers Wikipedia and Mondial datasets as well \cite{coffman2014}. Testing \textsf{Bsmooth} against these other datasets would be straightforward, if not for the cost of crowdsourcing the 50 queries (i.e., $\!50+\!$ terms) for each DB over multiple settings to simulate user feedback. Our work, nonetheless, establishes a first direction that can be easily incorporated into off-the-shelf IDR systems. 

The bottom line is that user feedback, even implicit, can greatly enhance the quality of disambiguation in IDR systems. The recommended strategy is LOGIT, tuned with a regularity factor ($H_D$) of .75, to balance the intrinsic and extrinsic evidence.  Algorithms~\ref{alg:profiling}--\ref{alg:smoothing} give the complete recipe.

\section*{References}
\bibliographystyle{elsarticle-num}
\bibliography{ijapprox}

\section{Appendix: Background review, proofs and fine-tuned parameters}

\subsection{Background Review on Bayesian Inference}\label{subsec:tutorial}
\noindent
We begin with a simple example 
to review relevant bayesian inference concepts. 
\vspace{-4pt}
\begin{myex}\label{ex:carnap}
An urn with balls of $k$ different colors (number of balls of each color is unknown), and at each trial a ball is extracted from the urn and then replaced for the next trial. For simplicity, let us fix $k=2$ (black and white balls). We are told the actual proportion of black balls can be either $\theta^\prime=0.67$ or $\theta^{\prime\prime}=0.50$. 
Now suppose we observe $n\!=\!3$ trials, $X_1=\CIRCLE,\, X_2=\Circle,\, X_3=\CIRCLE$. We want to predict the outcome of next trial, then need to estimate $P(X_4 \,|\, X_1, X_2, X_3)$.
$\Box$
\end{myex}

\noindent
The experiment $X_1, X_2, ..., X_n$ from Example \ref{ex:carnap} is a sequence of \emph{Bernoulli trials}, as $k=2$ and outcomes are independent (sampling with replacement). Each variable $X_j$ may take success ($X_j=1$, say, black) or failure ($X_j=0$, white), and they are said independent and identically distributed (iid). They follow the binomial distribution. 

\textbf{Binomial distribution}. Its probability mass function $f(s;\;  n,\, \theta)$ is given by Eq. \ref{eq:binomial}, where $0 \leq s \leq n$ is the number of successes in $n$ trials and $\theta$ is the guessed parameter.
\begin{eqnarray}
f(s\;|\; n,\, \theta) \;=\; \binom{n}{s}\; \theta^s\, (1-\theta)^{n-s}
\label{eq:binomial}
\vspace{-3pt}
\end{eqnarray}

\noindent
Since the order of outcomes is not relevant in Bernoulli trials \cite{murphy2012}, the vector $\vec{n}=(2, 1)$ is considered a \emph{sufficient statistics} for $S_3\!=\!(X_1\mapsto\CIRCLE,\, X_2\mapsto\Circle,\, X_3\mapsto\CIRCLE$). 
For $n\!=\!3$, a variable $X$ can then take any outcome from the sample space $\Omega \!=\! \{(0, 3),\, (1, 2),\, (2, 1),\, (3, 0)\}$. We write $X \sim Bin(n, \theta)$.  
By Eq. \ref{eq:binomial}, we can compute how likely the considered parameter hypotheses are given the observed data:\vspace{-3pt}\\

\indent $P(X\mapsto 2 \,|\, n, \theta^\prime\,) \;\,\,=\, \binom{3}{2}\, (0.67)^2\,(0.33)^1 = 0.444$\vspace{2pt}\\
\indent $P(X\mapsto 2 \,|\, n, \theta^{\prime\prime}\,) \,\,=\, \binom{3}{2}\, (0.50)^2\,(0.50)^1 = 0.375$

\textbf{Bayes' rule}. A baseline method we can apply to answer the question of Example \ref{ex:carnap} is the rule for bayesian update \cite{murphy2012}. It is given by Eq.~\ref{eq:bayes}, where the likelihood function $p( s \,|\,n, \theta)$ is as we have seen for the parameter hypotheses $\theta^\prime,\, \theta^{\prime\prime}$. 
\vspace{-10pt}
\begin{eqnarray}
p(\theta \,|\, s, n) \;=\; \frac{p( s \,|\,n, \theta) \; p(\theta)}{\sum_{\theta \in \Theta}{p( s \,|\, n, \theta) \; p(\theta)} }
\label{eq:bayes}
\vspace{-2pt}
\end{eqnarray}
Now let us assume a symmetric prior for each parameter hypothesis, i.e., $p(\theta)\!=\!1/2$ for all $\theta \in \Theta$, where $\Theta=\{0.67, 0.50\}$. 
Applying Eq.~\ref{eq:bayes} then gives us: $P(\theta^\prime |\, s, n) = .542,\, P(\theta^{\prime\prime} |\, s, n) = 0.458$. 
The evidence favors parameter hypothesis $\theta^\prime=0.67$. But the symmetric prior has effect on the \emph{posterior mean}: 
the expected value from our updated beliefs on the parameter hypotheses is $E(X_4 \,|\, X)=\sum_{\theta \in \Theta}{\theta\;p(\theta)} = 0.592$. We are led to estimate $P(X_4 \mapsto 1 \,|\, X \mapsto 2)=0.592$.

The focus of Eq~\ref{eq:bayes} is on updating our confidence on the parameter hypotheses, rather than on (directly) finding the optimal prediction for next trial $X_4$. This is because we are locked up in the discrete parameter space $\Theta=\{0.67, 0.50\}$. 
Next we relax this assumption from Example~\ref{ex:carnap} and move to estimation in a continuous parameter space $\theta \in [0, 1]$.

\textbf{Maximum Likelihood Estimation (MLE)}. 
In the case of a discrete number of possible outcomes, $k \in \mathbb N$, and continuous parameter space, $\theta \in [0, 1]$, 
MLE reduces to the simple formula of relative frequency (Eq.~\ref{eq:rf}) \cite{murphy2012}. 
\vspace{-5pt}
\begin{eqnarray}
\hat{\theta} = s/n \label{eq:rf}
\vspace{-4pt}
\end{eqnarray}
Now by Eq.~\ref{eq:rf} we can compute $P(X_4 \mapsto 1 \,|\, X \mapsto 2) \!=\! 2/3 \!\approx\! 0.67$. The future is predicted to be just like the past.

We see now that the gap between MLE and bayesian inference can be bridged. MLE is known to work well in the very long run if the data is assumed `true' but not when it is `effectively small'---i.e., when it is noisy and/or not very large. However, it can be smoothed in a form of bayesian inference where both the prior and its strength are input.

\textbf{Conjugate prior}. 
A prior $p(\theta)$ to be input in bayes' rule (Eq.~\ref{eq:bayes}) must only be in the 1-simplex $[0, 1]$. But for simpler computation and interpretation of results (how beliefs are changed), it would be convenient if the prior also had the same form as the likelihood function. That is, if we could set some hyper-parameters $\gamma_1, \gamma_2 \geq 0$ such that \mbox{$p(\theta) \propto \theta^{\gamma_1} \, (1-\theta)^{\gamma_2}$}. The posterior then takes the form: 
\begin{eqnarray}\label{eq:conjugate}
 \begin{aligned}
p(\theta \,|\, s, n) \,\propto\, p( s, n \,|\,\theta) \; p(\theta) = \theta^s (1 - \theta)^{n-s} \; \theta^{\gamma_1} (1 - \theta)^{\gamma_2}\\ = \theta^{s+\gamma_1} (1 - \theta)^{n-s+\gamma_2}
\end{aligned}
\vspace{-3pt}
\end{eqnarray}

\noindent
Now it is easier to see how bayesian inference updates the (hyper-)prior ($\gamma_1, \gamma_2$) based on the data ($s, n$). Intuitively, the hyper-parameters are called pseudo-counts that are added to the real counts ($s$ out of $n$) observed. Note that the strength of the prior in inference (its `effective sample size') shall then depend on the ratio $(\gamma_1 + \gamma_2) / n$. 

$\!$In sum, a class of prior distributions $p(\theta) \!\in \mathcal P$ is said a \emph{conjugate distribution} for a sampling model $p(y \,|\, \theta)$ if $p(\theta) \!\in \mathcal P$ implies that $p(\theta \,|\, y) \in \mathcal P$ for all $p(\theta) \!\in \mathcal P$ and data $y$.

\textbf{Beta distribution}. 
Given hyper-parameters that way, the beta distribution 
is a conjugate distribution for the binomial. If a beta prior is input to Bayes' rule with binomial likelihood, then we must also have some beta posterior. 
 
$\!$This distribution has probability density function $f(\theta;\, \alpha,\, \beta)$ given by Eq. \ref{eq:beta}, where $\alpha,\, \beta > 0$ are the hyper-parameters.
\begin{eqnarray}
f(\theta\;|\; \alpha,\, \beta) \;=\; \text{constant}\; \cdot\; \theta^{\alpha-1}\, (1-\theta)^{\beta-1}
\label{eq:beta}
\end{eqnarray}
In Eq. \ref{eq:beta}, \emph{constant} is $\frac{\Gamma(\alpha+\beta)}{\Gamma(\alpha)\, \Gamma(\beta)}$, where $\Gamma(.)$ is the gamma function. It is a normalization factor in Eq. \ref{eq:beta} to ensure that the probabilities integrate to one. 
The beta distribution is \emph{symmetric} if $\alpha=\beta$, and \emph{skewed} otherwise. In particular, we could simulate an unbiased prior by sampling $\theta \sim Beta(1, 1)$---as we see from Eq.~\ref{eq:beta}, the probability of sampling any $0 \leq \theta \leq 1$ is the same when $\alpha=\beta=1$. 

\textbf{Beta-binomial}. We now arrive at \emph{bayesian smoothing} \cite{murphy2012}. For the beta-binomial case, let $\alpha, \beta >0$ be the hyper-para\-meters and $s$ out of $n$ be the evidence data for bayesian inference. Then the posterior is $p(\theta\,|\, s, n) \propto Beta(\alpha+s,\, \beta + n -s)$. Now no sampling is needed. The mean of the beta distribution is $\bar{\theta}=\gamma_1/(\gamma_1+\gamma_2)$ for hyper-parameters $\gamma_1,\, \gamma_2$. We can get the estimate from the beta posterior mean (Eq.~\ref{eq:bb}). 
\vspace{-2pt}
\begin{eqnarray}
p(\theta\;|\; s, n) \;=\; \frac{\alpha + s}{\alpha+\beta+n}
\label{eq:bb}
\end{eqnarray}
\vspace{-1pt}
Note that Eq.~\ref{eq:bb} is a convex combination (weighted sum) of the beta prior mean and MLE. 

Setting $\alpha=\beta=1$ is the particular case that leads to Laplace's rule of succession $p(\theta\,|\, s, n) = (s + 1)\,/\,(n+2)$. 
For comparison with our previous attempts at estimation, it gives us $P(X_4 \mapsto 1 \,|\, X \mapsto 2) = 3/5 = 0.6$. 
We see that this (hyper-)prior is unbiased, but weaker than the MLE component as $n>\alpha+\beta$. 
It is called `add-one smoothing' \cite[p.~77]{murphy2012}. The pseudo-count is to ensure both outcomes $X_4 \mapsto 1$ and $X_4 \mapsto 0$ are possible no matter what the observations were like. That is, it avoids MLE's overfitting and possible sparsity (zero counts). MLE can be recovered by setting $\alpha=\beta=0$ to get back $P(X_4 \mapsto 1 \,|\, X \mapsto 2) = 2/3 \approx 0.67$. 

The sum $\alpha+\beta$ is the so-called concentration parameter---the prior's `effective sample size.' 
It is left open how to set this subjective parameter. 

\textbf{Dirichlet multinomial}. 
Instead of just the number of successes $s$, we will have a vector $\vec{n}=(n_1,\, ...,\, n_k)$ for the counts observed on each of $k \in \mathbb N$ possible outcomes given a parameter vector $\vec{\theta}=(\theta_1,\, ...,\, \theta_k)$. 
The multinomial distribution is a straightforward generalization of the binomial, and the Dirichlet distribution is a conjugate for the multinomial \cite[p.~47]{murphy2012}. 
Its probability density function $f(\vec{\theta};\; \vec{\alpha})$ has hyper-parameters $\alpha_i > 0$, for $1 \leq i \leq k$. 
The distribution is \emph{symmetric} when hyper-parameters $\alpha_i$ are all equal, and \emph{skewed} otherwise. Now if $k=2$ with $\vec{\alpha}=(1, 1)$, then $Dir(\vec{\alpha})\!=\!Beta(1, 1)$. 

Bayesian smoothing for the dirichlet-multinomial can be obtained from the dirichlet posterior mean as given by the equation below (referred as Eq.~\ref{eq:dm} in \S\ref{subsec:problem}), where $n_i$ and $\alpha_i$ are components of (resp.) the observed data and hyper-parameter vectors $\vec{n}$, $\vec{\alpha}$; and $n,\, \alpha$ are (resp.) the total data and the hyper-prior concentration parameter. 
\begin{eqnarray*}
p(\theta_i \,|\, \vec{n}) \!\!&=&\!\! \frac{n_i + \alpha_i}{n+\alpha}
\end{eqnarray*}
Eq.~\ref{eq:dm} is the basic formula we apply to the interactive data retrieval application described in this paper. The open, non-trivial problem is how to set the concentration parameter $\alpha$ (i.e., the strength of the prior). This paper presents empirical principles for that problem applied to the interactive data retrieval use case.

\subsection{Proof of Lemma ~\ref{lemma:convex}}\label{proof:lemma-convex}
\noindent
\emph{Let $\vec{x}(x_1, ..., x_k)$ and $\vec{y}(y_1, ..., y_k)$ be two scoring vectors and $t \neq u$ two of their bins. Then a convex combination $\vec{z}=\vec{x} \oplus \vec{y}$ with $z_t=z_u$ exists and is unique if, only if one of these mutually exclusive conditions hold:\vspace{3pt}\\
\indent \emph{(i)}$\;\;$ $(x_t - x_u)/(y_t - y_u) < 0$,\vspace{2pt}\\
\indent \emph{(ii)}$\;$ $(x_t - x_u) = 0$ but $(y_t - y_u) \neq 0$,\vspace{2pt}\\
\indent \emph{(iii)} $(x_t - x_u) \neq 0$ but $(y_t - y_u) = 0$}.

\begin{myproof}
Let $\vec{z}$ be defined $z_i=\omega_1\, x_i\, + \omega_2\,y_i$ for all bins $i$ and some $\omega_1, \omega_2 \geq 0$ such that $\omega_1+\omega_2=1$. Then, by Def.~\ref{def:convex}, we have $\vec{z}=\vec{x}\,\oplus\,\vec{y}$ and we can write $z_t=\omega_1\, x_t\, + \omega_2\,y_t$ and $z_u= \omega_1\, x_u\, + \omega_2\,y_u$. Now we are interested in existence and uniqueness conditions for $z_t=z_u$.

In fact, we can find $\omega_1, \omega_2 \geq 0$ to satisfy both $\omega_1 + \omega_2=1$ and $\omega_1\, x_t\, + \omega_2\,y_t = \omega_1\, x_u\, + \omega_2\,y_u$ over constants $x_t,y_t,x_u,y_u \geq 0$. That is, our problem reduces to the solution set of the following linear system for $\omega_1, \omega_2 \geq 0$:
\[\begin{cases}
\omega_1 + \omega_2=1\\
(x_t - x_u) \,\omega_1  +  (y_t - y_u) \,\omega_2 = 0
\end{cases}\]
We start with the `only if' statement. That is, we must show that if $(x_t - x_u) / (y_t - y_u) > 0$ or $(x_t - x_u)=(y_t - y_u) = 0$ then a solution either cannot exist or is not unique. 

In fact, suppose by contradiction that $(x_t - x_u) / (y_t - y_u) > 0$. By basic manipulation over the second equation we have $\omega_1 / \omega_2 = -(y_t - y_u) / (x_t - x_u)$. Since we have assumed $(x_t - x_u) / (y_t - y_u) > 0$, then it must be the case that $\omega_1 / \omega_2 < 0$. But $\omega_1, \omega_2 \geq 0$. \lightning. 
Now rather if $(x_t - x_u) = (y_t - y_u) = 0$, then the system is indeterminate---underconstrained to just $\omega_1 + \omega_2=1$. That is, a solution exists but is not unique. 

It remains to show the `if' statement. In fact, if $(x_t - x_u)=0$ but $(y_t - y_u) \neq 0$, then we easily infer $\omega_2=0,\, \omega_1=1$ and condition (ii) is shown. A symmetric argument works for the converse to cover condition (iii). Thus in both cases the solution exists and is unique. 

Now finally, recall that $\omega_1 / \omega_2 = -(y_t - y_u) / (x_t - x_u)$ and let $\lambda=-(y_t - y_u) / (x_t - x_u)$. Then $\omega_1 / \omega_2 = \lambda$ and by condition (i) we must have $\lambda > 0$. That is, we have both $\omega_1/\omega_2=\lambda$ and $\omega_1 + \omega_2=1$ for $\lambda > 0$---which leads to $\omega_2=1/(\lambda+1)$ and $\omega_1=\lambda/(\lambda+1)$, where $\lambda>0$ and depends only on the constants $x_t, x_u, y_t, y_u$. Therefore the solution exists and is unique. $\Box$
\end{myproof}

\vspace{-5pt}
\subsection{Proof of Theorem~\ref{thm:convex}}\label{proof:thm-convex}
\noindent
\emph{Let $\vec{x}=(x_1, ..., x_k)$ and $\vec{y}=(y_1, ..., y_k)$ be two scoring vectors with top bins $t \in \ceil*{\vec{x}}$ and $u \in \ceil*{\vec{y}}$. If these are disagreed top bins $t \asymp u$ of $\vec{x}$ and $\vec{y}$, then a convex combination $\vec{z}=\vec{x} \oplus \vec{y}$ with both $z_t=z_u$ exists and is unique}.

\begin{myproof} 
The proof is straightforward from Lemma~\ref{lemma:convex} and Defs.~\ref{def:top-bin}--\ref{def:disagreed-top-bins}. Note that if $t \asymp u$, then by Def.~\ref{def:disagreed-top-bins} we have $t \in \ceil*{\vec{x}}$ and $u \in \ceil*{\vec{y}}$ but either $t \notin \ceil*{\vec{y}}$ or $u \notin \ceil*{\vec{x}}$. We start with condition $t \notin \ceil*{\vec{y}}$. 

If $t \notin \ceil*{\vec{y}}$, then by Def.~\ref{def:top-bin} there must be some bin $s \neq t$ such that $y_t < y_s$. Also, since by assumption $u$ is a top bin of $\vec{y}$ we have $y_s \leq y_u$. Thus we must have $y_t < y_u$ and, therefore, $y_t - y_u < 0$. Now by Lemma~\ref{lemma:convex}.(i-ii), we know that $\vec{z}=\vec{x} \oplus \vec{y}$ with both $z_t=z_u$ exists and is unique if, only if $x_t - x_u \geq 0$. That is, when $x_t \geq x_u$. In fact, by assumption $t \in \ceil*{\vec{x}}$ then, again by Def.~\ref{def:top-bin}, that must be the case.

The argument for condition $u \notin \ceil*{\vec{x}}$ is analogous. $\Box$
\end{myproof}

\subsection{Empirical Validation of Fact~\ref{fact:logit}}\label{proof:fact}
\noindent
Tables~\ref{tab:linear-params}, \ref{tab:logit-params} present the empirical validation for Fact~\ref{fact:logit}. Some relevant observations are: 

\begin{itemize}
\item For each prior $\vec{x}_i$ the lack of fit for the linear profile (Table~\ref{tab:linear-params}) is always significantly larger than the lack of fit for the LOGIT profile (Table~\ref{tab:logit-params}).

\item The optimal parameters $\beta_0,\, \beta$ for the linear profile (Table~\ref{tab:linear-params}) are always close to the specific profile LINEAR, with $\beta_0=\beta=1$, as introduced in Remark~\ref{rmk:linear} and used in the evaluation of \S\ref{sec:eval}.

\item The lack of fit of LOGIT (Table~\ref{tab:logit-params}) is always very small. 
\end{itemize}

\begin{table}[t]
\caption{Optimal parameters $\beta_0,\, \beta$ for the linear profile, where $P=\{(0, 1),\, (1, 0),\, (.75, \bar{\omega}_2)\}$ and $\bar{\omega}_2$ varies row-wise.}
\label{tab:linear-params}
\centering
\begingroup\setlength{\fboxsep}{0.25pt}
\colorbox{gray!15}{%
\begin{tabular}{c | c | c | c | c}
\multicolumn{5}{c}{Linear profile }\\
\hline
prior$\!$ & $(D, \bar{\omega_2})$ & $\beta_0$ & $\beta$ & $\!$lack of fit$\!\!\!$\\
\hline
$\vec{x}_6$ & $(.75, .637)$ & $1.059$ & $0.880$ & $9.253$E-2$\!\!$\\
$\vec{x}_5$ & $(.75, .584)$ & $1.051$ & $0.897$ & $6.897$E-2$\!\!$\\
$\vec{x}_4$ & $(.75, .513)$ & $0.985$ & $1.048$ & $1.036$E-1$\!\!$\\
$\vec{x}_3$ & $(.75, .412)$ & $1.025$ & $0.950$ & $1.640$E-2$\!\!$\\
$\vec{x}_2$ & $(.75, .260)$ & $1.002$ & $0.997$ & $7.012$E-5$\!\!$\\
$\vec{x}_1$ & $(.75, .203)$ & $0.993$ & $1.014$ & $1.317$E-3$\!\!$\\
$\vec{x}_0$ & $(.75, .034)$ & $0.967$ & $1.066$ & $2.847$E-2$\!\!$\\
\end{tabular}
}\endgroup
\end{table}

\begin{table}[h!]
\caption{Optimal parameters $\beta_0,\, \beta$ for LOGIT$_3$, with $|P|=3$ where $P=\{(0, 1),\, (1, 0),\, (.75, \bar{\omega}_2)\}$ and $\bar{\omega}_2$ varies row-wise.}
\label{tab:logit-params}
\centering
\begingroup\setlength{\fboxsep}{0.25pt}
\colorbox{gray!15}{%
\begin{tabular}{c | c | c | c | c}
\multicolumn{5}{c}{LOGIT$_3$}\\
\hline
prior$\!$ & $(D, \bar{\omega_2})$ & $\beta_0$ & $\beta$ & $\!$lack of fit$\!\!\!$\\
\hline
$\vec{x}_6$ & $(.75, .637)$ & $244.559$ & $324.925$ & $6.989$E-11$\!\!$\\
$\vec{x}_5$ & $(.75, .584)$ & $253.127$ & $336.633$ & $1.813$E-11$\!\!$\\
$\vec{x}_4$ & $(.75, .513)$ & $261.633$ & $348.343$ & $2.653$E-11$\!\!$\\
$\vec{x}_3$ & $(.75, .412)$ & $181.671$ & $242.399$ & $7.216$E-10$\!\!$\\
$\vec{x}_2$ & $(.75, .260)$ & $162.432$ & $217.703$ & $4.839$E-10$\!\!$\\
$\vec{x}_1$ & $(.75, .203)$ & $050.270$ & $068.766$ & $9.709$E-13$\!\!$\\
$\vec{x}_0$ & $(.75, .034)$ & $138.507$ & $188.908$ & $3.469$E-11$\!\!$\\
\end{tabular}
}\endgroup
\end{table}

\vspace{-5pt}
\subsection{Final Tuned Parameters for LOGIT}\label{subsec:tuned-logit}
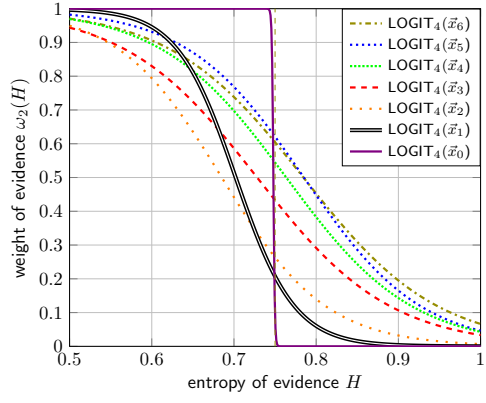
\begin{figure}[h]\centering
\begin{tikzpicture}[font=\sffamily,scale=0.7]
\begin{axis}
[
xmajorgrids = true,
ymajorgrids = true,
height=6.15cm,
width=7.1cm,
x label style = { yshift=0.1cm },
y label style = { yshift=-0.3cm },
xlabel= \textsf{entropy of evidence $H$},ylabel= \textsf{weight of evidence $\omega_2(H)$},xtick={0.5,0.6,...,1.1},ytick={0,0.1,...,1.1},
xmin=0.5,xmax=1.0,ymin=0,ymax=1,tick,samples=500,
domain=0.5:1.1,restrict y to domain =0:1.15,scale=1.4,legend style={at={(1,1)}}
]
 \addplot[very thick,domain=0.5:1,dashdotted,variable=\x,olive!100] plot ({\x},{1/(1+exp(-9.628+12.272*\x))});
 \addplot[very thick,domain=0.5:1,dotted,variable=\x,blue!100] plot ({\x},{1/(1+exp(-11.065+14.094*\x))});
 \addplot[very thick,domain=0.5:1,densely dotted,variable=\x,green!100] plot ({\x},{1/(1+exp(-10.018+13.119*\x))});
 \addplot[very thick,domain=0.5:1,dashed,variable=\x,red!100] plot ({\x},{1/(1+exp(-9.009+12.372*\x))});
 \addplot[very thick,domain=0.5:1,loosely dotted,variable=\x,orange!100] plot ({\x},{1/(1+exp(-10.854+15.841*\x))});
 \addplot[thick,domain=0.5:1,double,variable=\x,black] plot ({\x},{1/(1+exp(-19.654+27.994*\x))});
 \addplot[very thick,domain=0.5:1,solid,variable=\x,violet!100] plot ({\x},{1/(1+exp(-842.589+1126.518*\x))});
\addplot +[color=olive!100, mark=none, dashed] coordinates {(0.75, 0) (0.75, 1)};
\legend{ \footnotesize{LOGIT$_4(\vec{x}_6$)}, \footnotesize{LOGIT$_4(\vec{x}_5$)}, \footnotesize{LOGIT$_4(\vec{x}_4$)}, \footnotesize{LOGIT$_4(\vec{x}_3$)}, \footnotesize{LOGIT$_4(\vec{x}_2$)}, \footnotesize{LOGIT$_4(\vec{x}_1$)}, \footnotesize{LOGIT$_4(\vec{x}_0$)}}
\end{axis}
\end{tikzpicture}
\caption{Tuned curve profiles for LOGIT$_4$, with $|P|=4$ where $P^\prime=\{(0, 1),\, (1, 0),\, (.75, \bar{\omega}_2), (.80, \bar{\omega}_2^\prime)\}$. The setting for $\vec{x}_1$ (double line) is the one used in our evaluation in \S\ref{sec:eval}.
}
\label{fig:logit-priors}
\vspace{-20pt}
\end{figure}

\noindent
Table~\ref{tab:tuned-logit} presents the final tuned parameters for LOGIT$_4$, with $|P|=4$ where $P^\prime=\{(0, 1),\, (1, 0),$ $(.75, \bar{\omega}_2), (.75, \bar{\omega}_2^\prime)\}$. 
Fig.~\ref{fig:logit-priors} shows the resulting curve profiles. We observe: 

\begin{itemize}
\item The nearly constant $\bar{\omega}_2$ for prior $\vec{x}_0$ near $H_D=.75$ in Fig.~\ref{fig:priors} leaves no room for smoothness in its corresponding profile in Fig.~\ref{fig:logit-priors}, whose form is step-like. 

\item For the other priors we can have smoother profiles with a slope that is roughly the same. 
\end{itemize}

\begin{table}[t!]
\caption{Tuned parameters $\beta_0,\, \beta$ for the LOGIT$_4$ profile, with $|P|=4$ where $P^\prime=\{(0, 1),\, (1, 0),\, (.75, \bar{\omega}_2), (.80, \bar{\omega}_2^\prime)\}$. The setting for $\vec{x}_1$ is the one used in our evaluation in \S\ref{sec:eval}.}
\label{tab:tuned-logit}
\centering
\begingroup\setlength{\fboxsep}{0.25pt}
\colorbox{gray!15}{%
\begin{tabular}{c | c | c | c | c}
\multicolumn{5}{c}{LOGIT$_4$ --- FINAL TUNED PARAMETERS}\\
\hline
prior$\!$ & $(D, \bar{\omega_2})$ & $\beta_0$ & $\beta$ & $\!$lack of fit$\!\!\!$\\
\hline
$\vec{x}_6$ & $(.75, .637)$ & $09.628$ & $12.272$ & $1.597$E-2$\!\!$\\
$\vec{x}_5$ & $(.75, .584)$ & $11.065$ & $14.094$ & $4.972$E-3$\!\!$\\
$\vec{x}_4$ & $(.75, .513)$ & $10.018$ & $13.119$ & $4.099$E-3$\!\!$\\
$\vec{x}_3$ & $(.75, .412)$ & $09.009$ & $12.372$ & $2.131$E-3$\!\!$\\
$\vec{x}_2$ & $(.75, .260)$ & $10.854$ & $15.841$ & $5.074$E-5$\!\!$\\
\rowcolor{gray!.20}$\vec{x}_1$ & $(.75, .203)$ & $19.654$ & $27.994$ & $1.110$E-7$\!\!$\\
$\vec{x}_0$ & $(.75, .034)$ & $842.59$ & $1126.5$ & $4.84$E-10$\!\!$\\
\end{tabular}
}\endgroup
\end{table}

\end{document}